\documentclass{aa}  
\renewcommand\makeLineNumber{}

\usepackage{newtxtext,newtxmath}

\usepackage[T1]{fontenc}
\usepackage{ae,aecompl}
\usepackage{comment}

\usepackage{aas_macros}
\usepackage[hidelinks]{hyperref}

\usepackage{graphicx}	
\usepackage[export]{adjustbox}

\usepackage{xcolor}

\def\gsim{\;\raise0.3ex\hbox{$>$\kern-0.75em\raise-1.1ex\hbox{$\sim$}}\;}

\def\lsim{\;\raise0.3ex\hbox{$<$\kern-0.75em\raise-1.1ex\hbox{$\sim$}}\;}
\def\gsim{\;\raise0.3ex\hbox{$>$\kern-0.75em\raise-1.1ex\hbox{$\sim$}}\;}
\begin{document}

\title{Perseus cluster in its X-ray entirety with SRG/eROSITA}
\subtitle{Merger and radio-uroboroses}

\author{
Eugene~Churazov \inst{1,2} 
\and
Ildar~I.~Khabibullin \inst{3,2,1} 
\and
Natalya~Lyskova \inst{2}
\and
Rashid~A.~Sunyaev \inst{2,1} 
\and Klaus~Dolag \inst{3,1}
}

\institute{
Max Planck Institute for Astrophysics, Karl-Schwarzschild-Str. 1, D-85741 Garching, Germany 
\and 
Space Research Institute (IKI), Profsoyuznaya 84/32, Moscow 117997, Russia
\and
Universitäts-Sternwarte, Fakultät für Physik, Ludwig-Maximilians-Universität München, Scheinerstr.1, 81679 München, Germany
}

\abstract{
The Perseus cluster (Abell 426) is a nearby massive galaxy cluster that spans several degrees.   We combined SRG/eROSITA, XMM-Newton, and Chandra data to get a complete coverage of this cluster in X-rays up to $R_{\rm 200c}$ and beyond, although at the largest radii, spatial non-uniformities of the X-ray sky background and foreground dominate. While the Perseus central part represents a canonical cool-core structure with clear signs of AGN Feedback, the outskirts, in turn, serve as a convincing example of a merger-perturbed system.  X-ray data suggest that IC310 is the main galaxy of a subcluster that merges with Perseus over the past $\sim 4\,{\rm Gyr}$. Overall, this configuration resembles the merger between the Coma cluster and the NGC4839 group. It is statistically more likely to find a merging group near the apocenter of its orbit. Therefore, it is not surprising that IC310 in Perseus has a relatively small velocity relative to the main cluster, similarly to NGC4839 in Coma.

Perseus also hosts a high-velocity radio galaxy, NGC1265 (line-of-sight velocity is almost twice the virial velocity of the main cluster), which is known for its spectacular radio tail.  Unless this galaxy has been accelerated by a time-variable potential associated with the merger, it has to move almost along the line of sight through the entire cluster, which would be a rare, but not a truly exceptional configuration. Both galaxies, IC310 and NGC1265, have remarkable radio tails with sharp bends that are reminiscent of a "snake biting its tail".   We speculate that these curious shapes are natural consequences of their (different) orbits in Perseus. For IC310, the proximity to the apocenter and the reversal of its radial velocity might play a role. For NGC1265, the nearly line-of-sight motion coupled with the gas motions in the merging system might be important.
}

\titlerunning{Perseus clusters in X-rays}

  
\maketitle

\section{Introduction}
\label{sec:intro}

The Perseus cluster (Abell 426) is the X-ray brightest and one of the best-studied clusters. Its core properties were used to develop the concept of "AGN feedback" \citep[e.g.,][]{1993MNRAS.264L..25B,2000A&A...356..788C,2003MNRAS.344L..43F,2015MNRAS.450.4184Z}. In the core and at larger scales, X-ray images show signs of perturbations, including a sequence of sloshing patterns located at different distances from the center  \citep[e.g.,][]{1981ApJ...248...55B,1992A&A...256L..11S,1998MNRAS.300..837E,2003ApJ...590..225C,2011MNRAS.418.2154F,2012ApJ...757..182S,2014ApJ...782...38T,2020MNRAS.498L.130Z,2020A&A...633A..42S,2022ApJ...929...37W,2021A&A...652A.147Z}, \cite[see, also ][for reviews on merger-induced perturbations]{2007PhR...443....1M,2016JPlPh..82c5301Z}.

The complications associated with the X-ray analysis of the Perseus cluster are caused by a) proximity to the Galactic Plane (hence, significant photoelectric absorption) and the large angular size, which complicates mapping the entire cluster with telescopes having a relatively small field of view.  The latter problem can be overcome with all-sky surveys. In this study, we report SRG/eROSITA \citep{2021A&A...656A.132S,2021A&A...647A...1P} observations of the Perseus cluster up to its virial radius. We combine eROSITA data with the publicly available Chandra and XMM-Newton data to achieve high angular resolution in the core of the cluster and 
a uniform coverage in the outskirts.

In this study, we adopted the following basic properties of the Perseus cluster: $z_{\rm cl}=0.01767$ \citep{2016Natur.535..117H}.  For $h=0.7$, this redshift translates to the angular diameter distance of 74.1~Mpc; $1'=21.5\,{\rm kpc}$. We adopt\footnote{See also, \cite{2020A&A...640A..30M} for another set of parameters which imply a larger mass of the Perseus cluster.} $R_{\rm 200c} = 1.79\,{\rm Mpc}$ and $M_{\rm 200c}=6.65 \times 10^{14}\,M_\odot$ \citep{2011Sci...331.1576S} and the corresponding circular velocity $V_{\rm 200c}=\left ( GM/r\right )^{1/2}=1263\,{\rm km\,s^{-1}}$. We also assume that $R_{\rm 200c}$ can be used in place of the virial radius (a parameter that enters the mass model). 

\section{X-ray data}
\label{s:x-rays}

\begin{figure*}
\includegraphics[angle=0,clip,width=2\columnwidth]{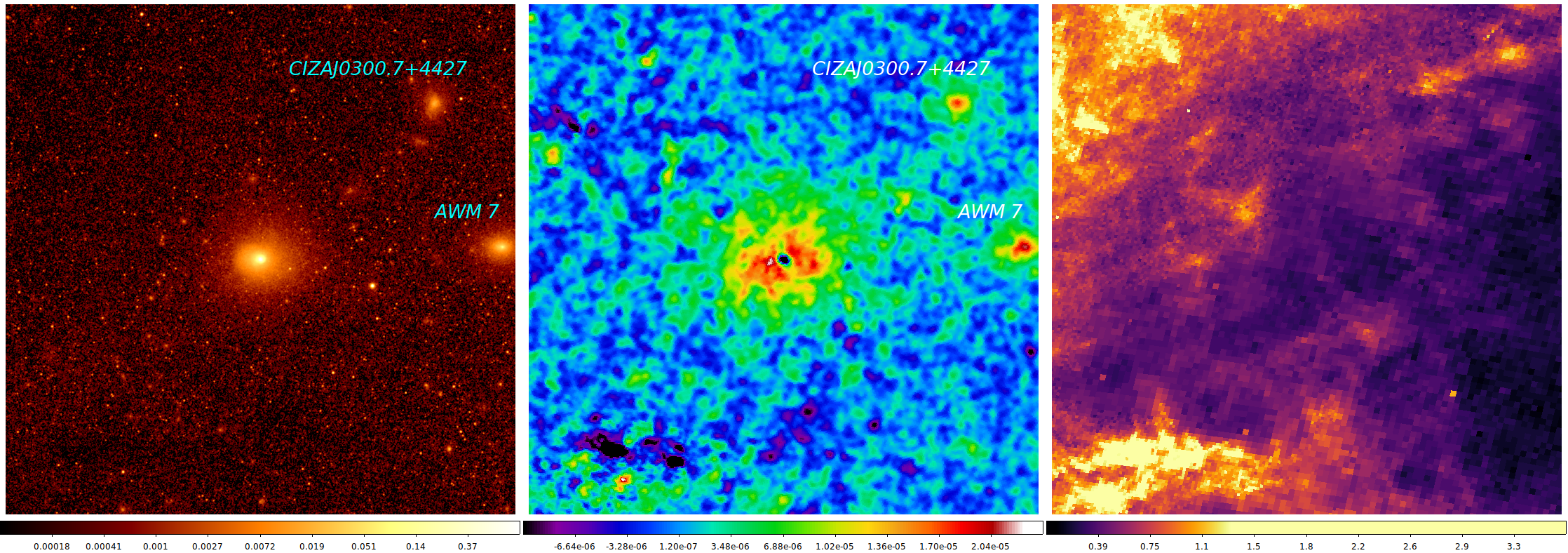}
\caption{10x10 degrees patch of the sky centered at the Perseus cluster in equatorial coordinates. The estimated turnaround radius of the Perseus cluster is $\sim 7^\circ$, i.e., the entire image falls within the turnaround radius. 
Left: eROSITA 0.4-2.3~keV X-ray image; middle: $y$-map from PLANCK based on PR4 maps \citep[see][]{2023MNRAS.526.5682C}; right: dust extinction map \citep[based on][]{2019ApJ...887...93G}. 
Two X-ray and SZ-bright objects on the right are the galaxy clusters AWM7/2A0251+413  ($z\sim 0.0172$) and CIZAJ0300.7+4427 ($z\sim 0.030$).  
These two clusters possibly trace the topology of LSS filaments, in particular, AWM7, which belongs to the Perseus-Pisces supercluster \citep[e.g.,][]{2021A&A...651A..16B}. CIZAJ0300.7+4427 belongs to another (more distant) filament, which is also seen in the distribution of galaxies (see Sect.~\ref{s:2mass}).  
The correspondence between X-ray and Y maps is very good, with the visible SZ signal being less peaked at the center but, as expected, extending to large radii. The "dip" in the very core of the $y$-map is due to contamination by the radio galaxy NGC1275 (Perseus A) and its immediate vicinity.   The right figure shows the complexity of the foreground distribution of the Milky Way gas (as traced by dust) in the direction of Perseus. The map is plotted in units of $A_V$ extinction. 
} 
\label{f:ero_sz_dust}
\end{figure*}

\begin{figure*}
\centering
\includegraphics[angle=0,clip,trim=0.5cm 5.5cm 1cm 2.5cm,clip,width=0.99\columnwidth]{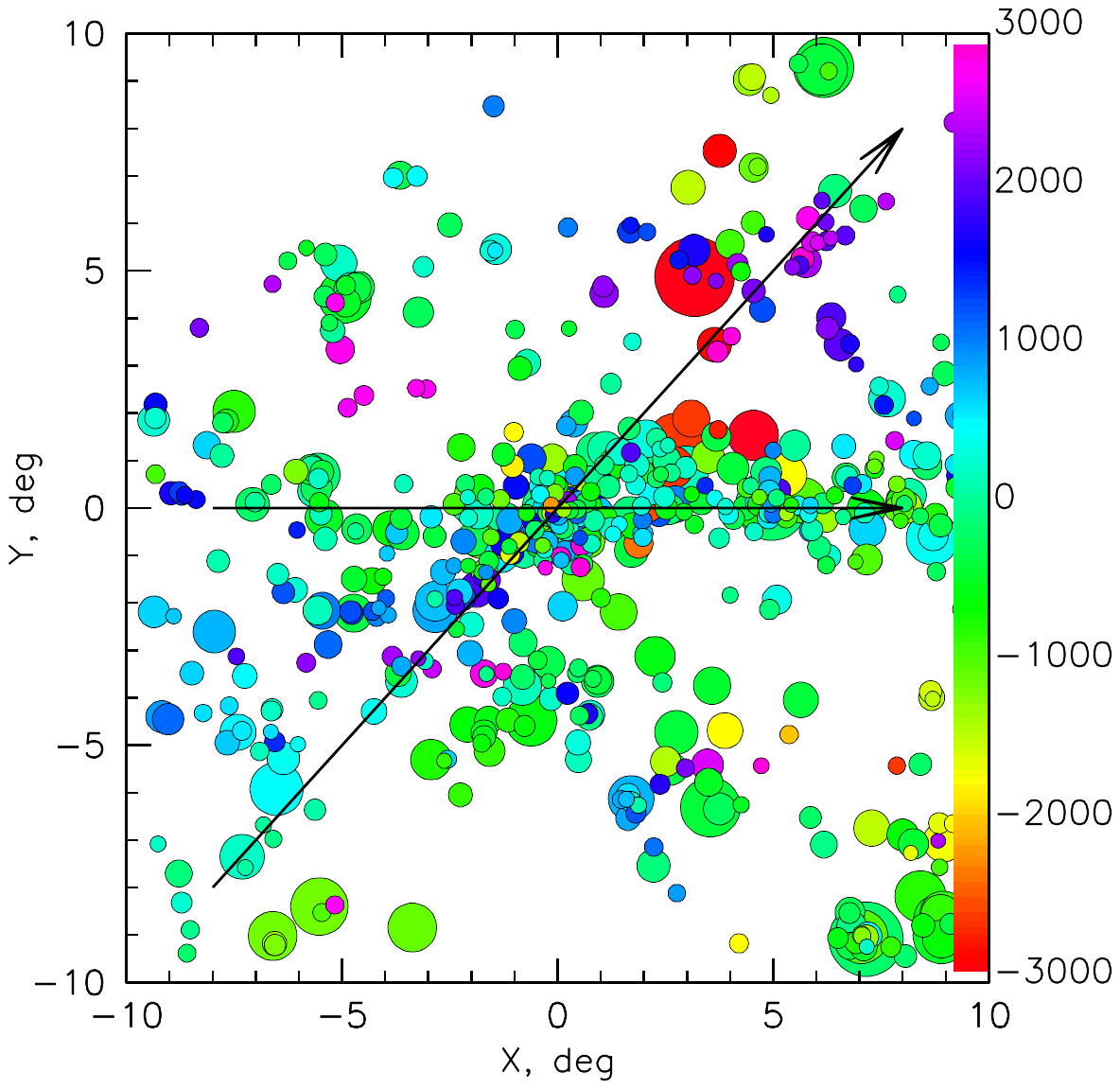}
\includegraphics[angle=0,clip,trim=0.5cm 5.5cm 1cm 2.5cm,clip,width=0.99\columnwidth]{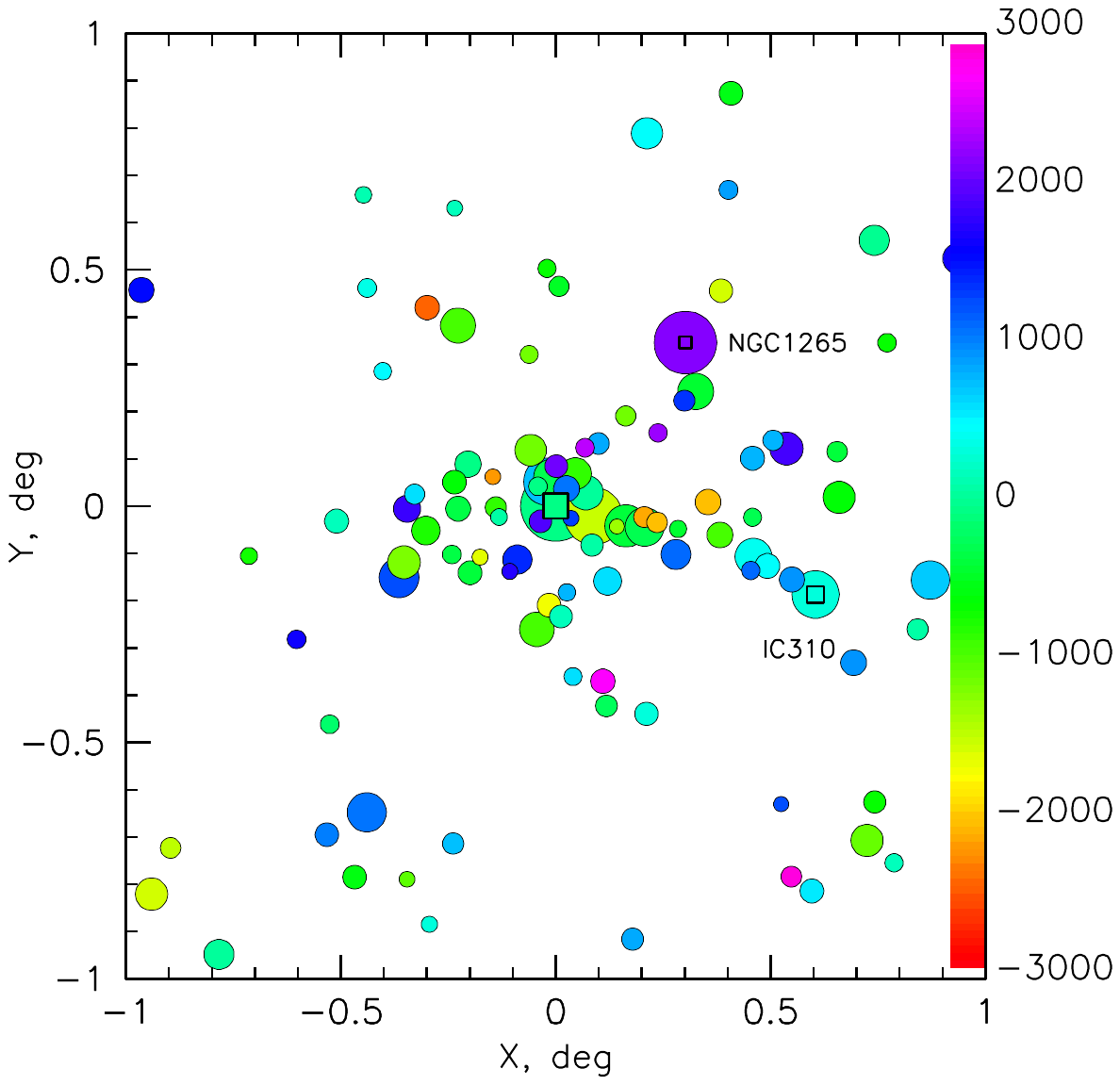}
\caption{Positions and recession velocities of 2MASS galaxies in the vicinity of NGC1275 with velocities in the range $\pm 3000\,{\rm km\,s^{-1}}$ relative to NGC1275.  The size of the symbols reflects the K magnitudes of a galaxy (brighter - larger), and the color characterizes the line-of-sight velocity relative to NGC1275 in ${\rm km\,s^{-1}}$ (see the color bar on the right side of the plot). 
{Left:} 20x20 degrees images showing large-scale filaments. Two of them are marked with black lines. { Right:}  2x2 degrees version of the same plot. NGC1275, NGC1265, and IC310 are marked with squares. A clear elongation of galaxies with small line-of-sight velocities in the East-West direction suggests that IC310 is part of a "filament" that spans at least a few degrees.
In turn, NGC1265 might be related to another "diagonal" (bottom-left to top-right) filament that has several galaxies with larger recession velocities (blue/purple circles in the plot). 
}
\label{f:vel_map}
\end{figure*}

For X-ray analysis, we use the data accumulated by the eROSITA \citep{2021A&A...647A...1P} telescope onboard the SRG~ X-ray observatory \citep{2021A&A...656A.132S} in the course of its all-sky survey.
Initial reduction and processing of the data were performed using standard routines of the \texttt{eSASS} software \citep{2018SPIE10699E..5GB,2021A&A...647A...1P}, while the imaging analysis was carried out with the detector background models, vignetting, point spread function (PSF) and spectral response function calibrations built upon the standard ones via slight modifications motivated by results of calibration and performance verification observations \citep[e.g.][]{2021A&A...651A..41C,2023MNRAS.521.5536K}. These modifications are intended to reproduce best the effective (i.e., averaged over multiple source transits of the telescope's Field-of-View during scanning mode observations) characteristics, including the averaged energy-dependent effective area and extended wings of the PSF (see Appendix\ref{a:obsids} for details).  Given the very bright central core of the Perseus cluster, the effect of the stray light scattering, included in the extended PSF model, might be particularly relevant for the radial profile measurement beyond $R_{200}$ of the cluster. At these radii, the emission of the cluster becomes comparable or smaller than the sum of the astrophysical and instrumental background.
The eROSITA's detector background is well measured and was very stable during most of the sky survey after exclusion of periods of enhanced Solar activity  \citep[see, e.g.,][]{2021A&A...647A...1P,2021A&A...656A.132S,2021SPIE11444E..1OF}.

For XMM-Newton and Chandra, we used a subset of observations accumulated over the lifetime of these missions. The lists of observations are given in the Appendix~\ref{a:obsids}.

\subsection{Large-scale image}
\label{s:large}

A 10x10 degrees patch of the sky centered at the Perseus cluster is shown in Fig.~\ref{f:ero_sz_dust}. The eROSITA 0.4-2.3 keV image illustrates the location of the Perseus cluster with respect to the large-scale structure traced by massive halos. In particular, two bright clusters AWM7/2A0251+413 ($z\sim 0.0172$) and CIZAJ0300.7+4427 ($z\sim 0.030$) are clearly visible. AWM7 together with Perseus belongs to the Perseus-Pisces supercluster \citep[e.g.,][]{2021A&A...651A..16B} and is well aligned with a major filament going through Perseus in the East-West direction. CIZAJ0300.7+4427 has a larger recession velocity, and the direct connection to Perseus is less certain. The same clusters are also visible in the PLANCK $y$-parameter map (Fig.~\ref{f:ero_sz_dust}, middle panel), where $y$ is proportional to the line-of-sight integrated electron pressure and can be derived from distortions of the Cosmic Microwave Background \citep[see][where the formulae describing these distortions were found]{1969Ap&SS...4..301Z}.   

The location of the Perseus cluster close to the Galactic Plane introduces some biases in the X-ray and $y$-maps. The $A_V$ extinction derived from the Bayestar data \citep{2019ApJ...887...93G} varies from 0.1 to 3.6 across the image. As a result, significant variations of photoelectric absorption modulate the X-ray image, and the dust-correlated structures are also visible in the $y$-map (see, e.g., the bottom-left corner of the images). Assuming that $A_V$ can be converted to equivalent hydrogen column density as $N_H\approx 2.2\times 10^{21} A_V$ \citep[e.g.,][]{1995A&A...293..889P,2009MNRAS.400.2050G} and considering the contribution of atomic gas (traced by HI, $N_{\rm H}\sim 1.3\times 10^{21}\,{\rm cm^2}$), we can expect a modulation of the Perseus X-ray flux in the 0.4-2.3 keV up to $\sim 30\%$ within a circle of 1.4 degrees. This estimate is done for the APEC spectrum with a temperature of 5~keV. The total attenuation factor of the cluster flux (relative to $N_{\rm H}=0$) is about 0.55. We further assess the impact of this modulation in Appendix~\ref{a:nh}.

\subsection{2MASS}
\label{s:2mass}
It is useful to compare the X-ray images with the distribution of galaxies with redshifts comparable to that of the cluster. To this end, we used the 2MASS Redshift Survey \citep{2012ApJS..199...26H} and selected galaxies with line-of-sight recession velocity in the range from $\varv_{\rm cl}-3000$ to $\varv_{\rm cl}+3000\,{\rm km\,s^{-1}}$. 
The distribution of these galaxies is shown in Fig.~\ref{f:vel_map}. The left (20x20 degrees) image shows clearly the famous large-scale filaments in the vicinity of Perseus, which are part of the Perseus-Pisces supercluster. The two most prominent filaments, relevant for this study, are marked with black lines. One of them is nearly horizontal in equatorial coordinates and is dominated by galaxies with redshifts similar to the main cluster. Another one is diagonal in this plot, and its upper end is dominated by galaxies with larger recession velocities (blue colors). The right image (2x2 degrees) is the zoomed version of the same plot. The chain of galaxies between NGC1275 and IC310 appears to align with the former filament, while NGC1265 might be part of the latter.

\subsection{X-ray composite image}
\label{s:ximage}
Focusing on the X-ray emission inside the Perseus virial radius, we would like to complement the eROSITA coverage of the entire field with the higher angular resolution of Chandra and XMM-Newton in the Perseus central region. 
We first generated separate detector-background-subtracted, exposure and vignetting-corrected X-ray images for eROSITA (0.4-2.3~keV; 4" pixels), XMM-Newton (0.5-3.5~keV; 2" pixels), and Chandra (0.5-3.5~keV, 1" pixels). These images have been combined in two steps. In the first step, the mean surface brightness in a circle centered at NGC1275 has been calculated, and the normalization of images was adjusted to compensate for slight differences in the energy bands and effective areas. The resulting images were rebinned to 2" resolution and co-added using a position-dependent weighting factor. This factor favors the Chandra images in the core and the eROSITA image at the largest distances from NGC1275. The resulting 2"-resolution, $3\times 3$ degrees image $I_X$ is shown in Fig.~\ref{f:cxee}. The inset shows the central 1\% of the same image, where Chandra's resolution captures familiar structures of the X-ray emission in the Perseus core.  

\subsection{Departures from a symmetric model}
\begin{figure*}
\sidecaption
\includegraphics[angle=0,trim=1.5cm 5cm 1cm 5.3cm,clip,width=12cm]{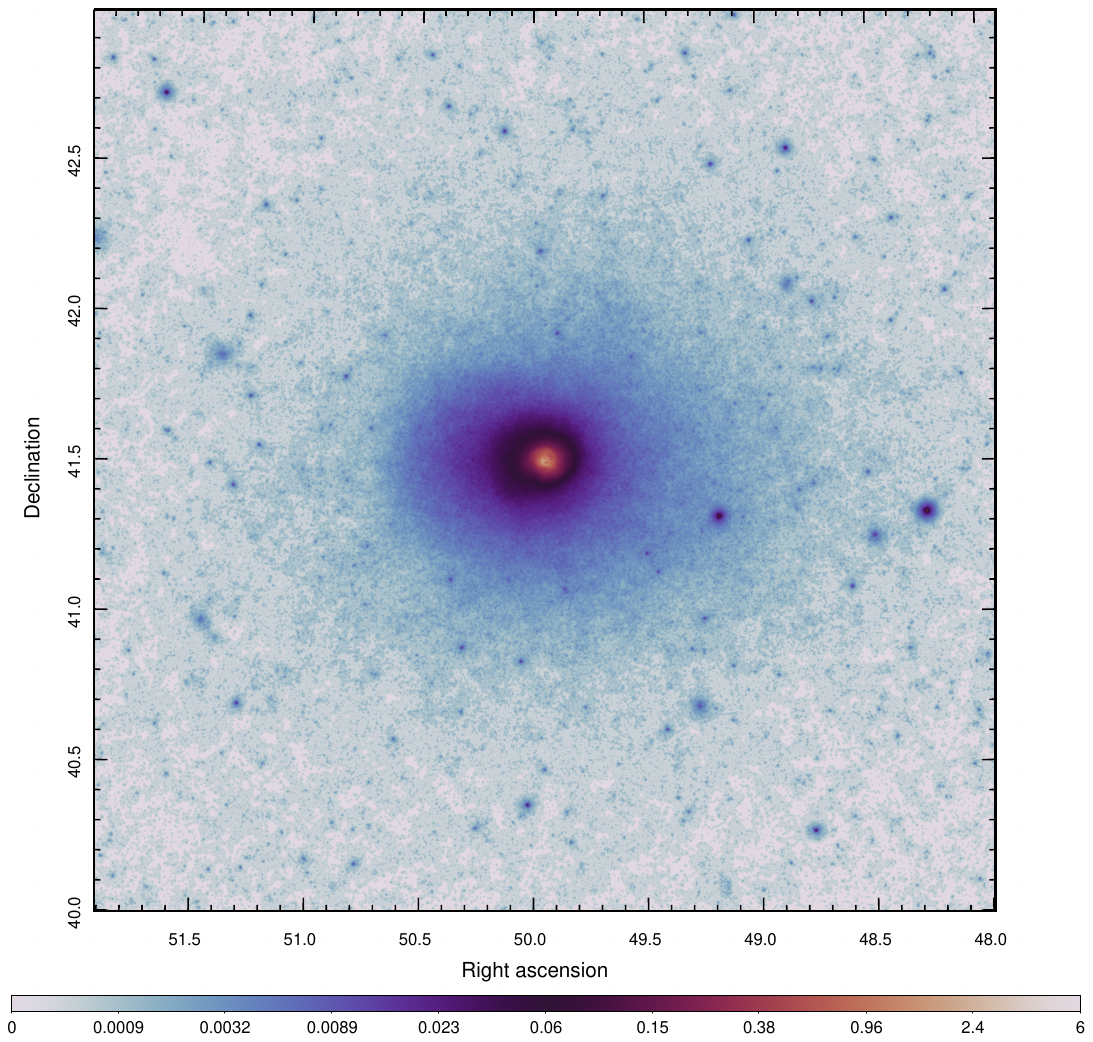}
\llap{\raisebox{1cm} {\includegraphics[angle=0,trim=1.8cm 5cm 1.8cm 5cm,frame,width=0.5\columnwidth]{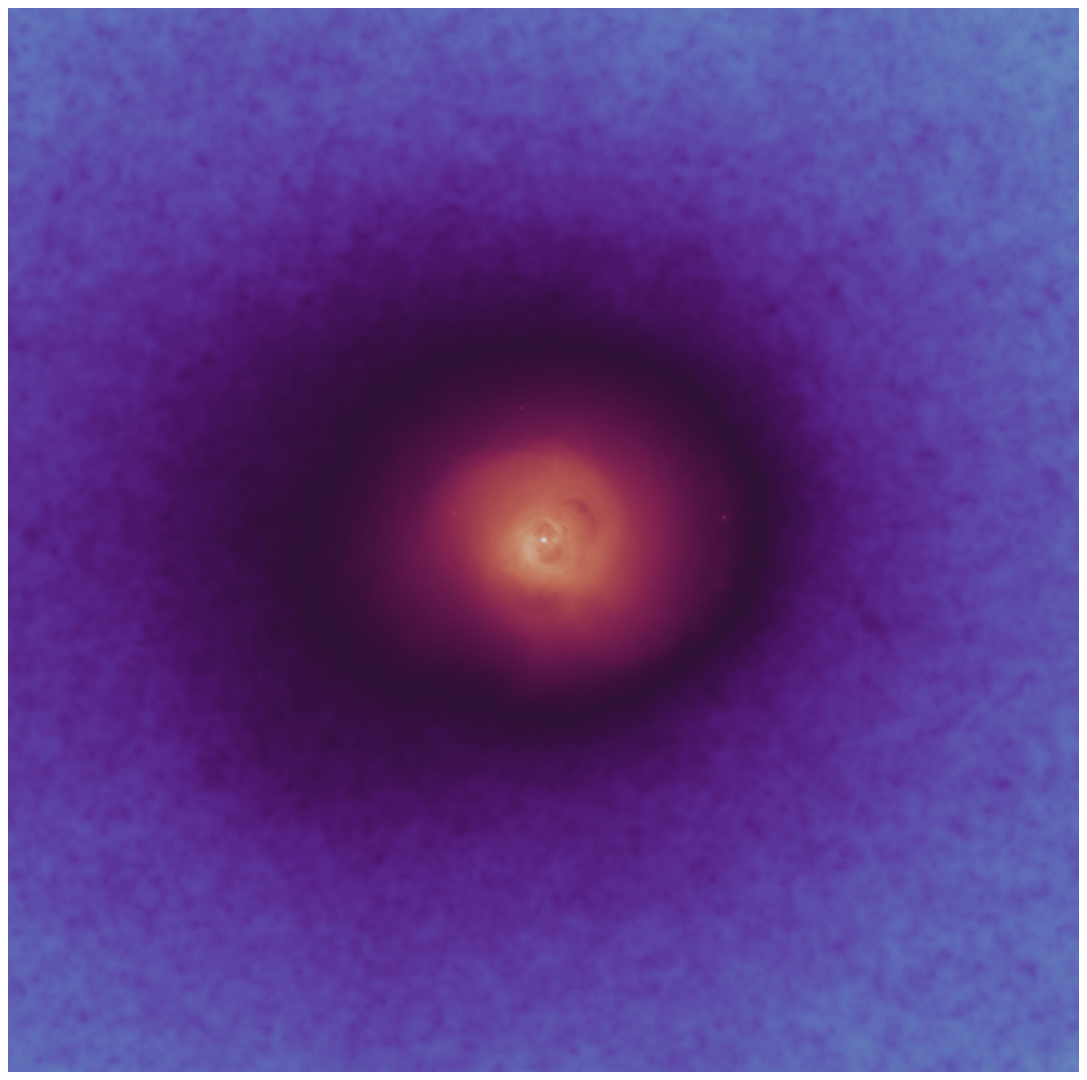}} }
\caption{Composite X-ray image of the Perseus cluster ($3\times 3$ degrees, 2" pixels) obtained by co-adding eROSITA all-sky data and pointed observations of Chandra and XMM-Newton. The images of individual telescopes were renormalized to have the same X-ray flux within the central $3'$ circle as in the eROSITA 0.4-2.3 keV band. Radial weights have been applied to ensure that the sharpest Chandra images dominate within a central $3'$. The inset is a zoomed version of the same image showing the central 1\% of the image (in terms of the solid angle, i.e., $0.3\times 0.3$ degrees) and the canonical cool-core design with an AGN (NGC1275) and X-ray cavities inflated by the AGN.} 
\label{f:cxee}
\end{figure*}

The known elongation of the Perseus X-ray emission along the East-West direction is already visible in Fig.~\ref{f:cxee}. To emphasize further the departures from spherical symmetry, we fitted the simplest beta model $I_{\rm model}=I_0/\left [ 1+\left ( r/r_c \right)^2\right ]^{3\beta-1/2}+B_{\rm s}$ to the radial X-ray surface brightness profile. Here, $B_{\rm s}$ is the spatially constant sky X-ray background. The brightest compact sources in the field were excised prior to fitting the model. The profile and the best-fitting beta model are shown in the Appendix~\ref{a:radial}. The dynamic range covered by the profile from the inner $\sim 2''$ to $\sim 100-200'$ approaches a factor of $10^5$. Overall, the beta model gives a reasonable (for our purposes) approximation of the Perseus X-ray surface brightness. We note in passing that at radii in the range 1-3 degrees, a stray light can give a non-negligible or even dominant contribution to the surface brightness profile. However, to reveal asymmetric features in the outskirts of the Perseus cluster, this is not required as long as the flux from the cluster core dominates the stray light and, therefore, the model remains symmetric, even though the amplitude of deviations can be affected. 

\begin{figure*}
%\centering
\sidecaption
\includegraphics[angle=0,clip,trim=1.5cm 5cm 3cm 5.3cm,width=12cm]{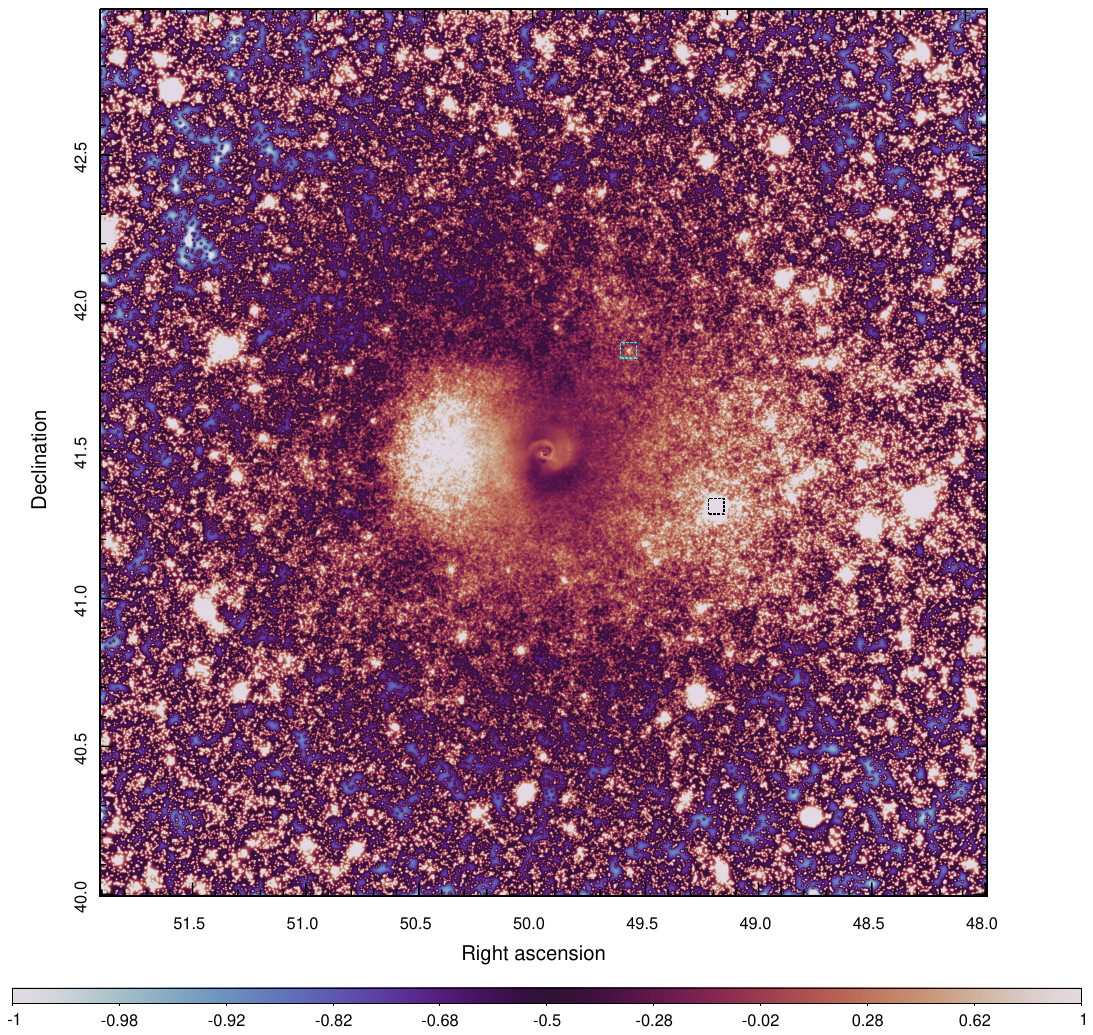}
\caption{The same image as in Fig.~\ref{f:cxee} after division by the best-fitting radial profile shown in Fig.~\ref{f:radial}, specifically,  $\left (I_X/I_{\rm model}-1 \right )$, where the model is the sum of the best-fitting $\beta$-model and a constant sky background). In addition to the rich substructure in the core, this image shows clear signatures of an ongoing merger, including elongation in the East-West direction and sloshing. Two boxes show the positions of two prominent radio galaxies, IC310 (black) and NGC1265 (cyan), respectively. (See also Fig.~\ref{f:divc_200} for a smoothed version of the image that emphasizes large-scale structures.) } 
\label{f:divc}
\end{figure*}

Fig.~\ref{f:divc} shows the X-ray image divided by the model,  specifically, $\left (I_X/I_{\rm model}-1\right)$. 
Once again, the core shows the familiar pattern of X-ray cavities and several spiral-like structures, while on larger scales, the dominant features are two large regions of excess emission on both sides of the core, which are roughly aligned with the East-West direction. Many of these structures (or parts of them) have already been seen in previous X-ray observations \citep[e.g.,][]{1981ApJ...248...55B,1992A&A...256L..11S,1998MNRAS.300..837E,2003ApJ...590..225C,2011MNRAS.418.2154F,2012ApJ...757..182S,2014ApJ...782...38T,2022ApJ...929...37W,2021A&A...652A.147Z}, but the complete and uniform coverage of eROSITA reveals this pattern in its entirety.

\begin{figure*}
\centering
\includegraphics[angle=0,clip,width=2\columnwidth]{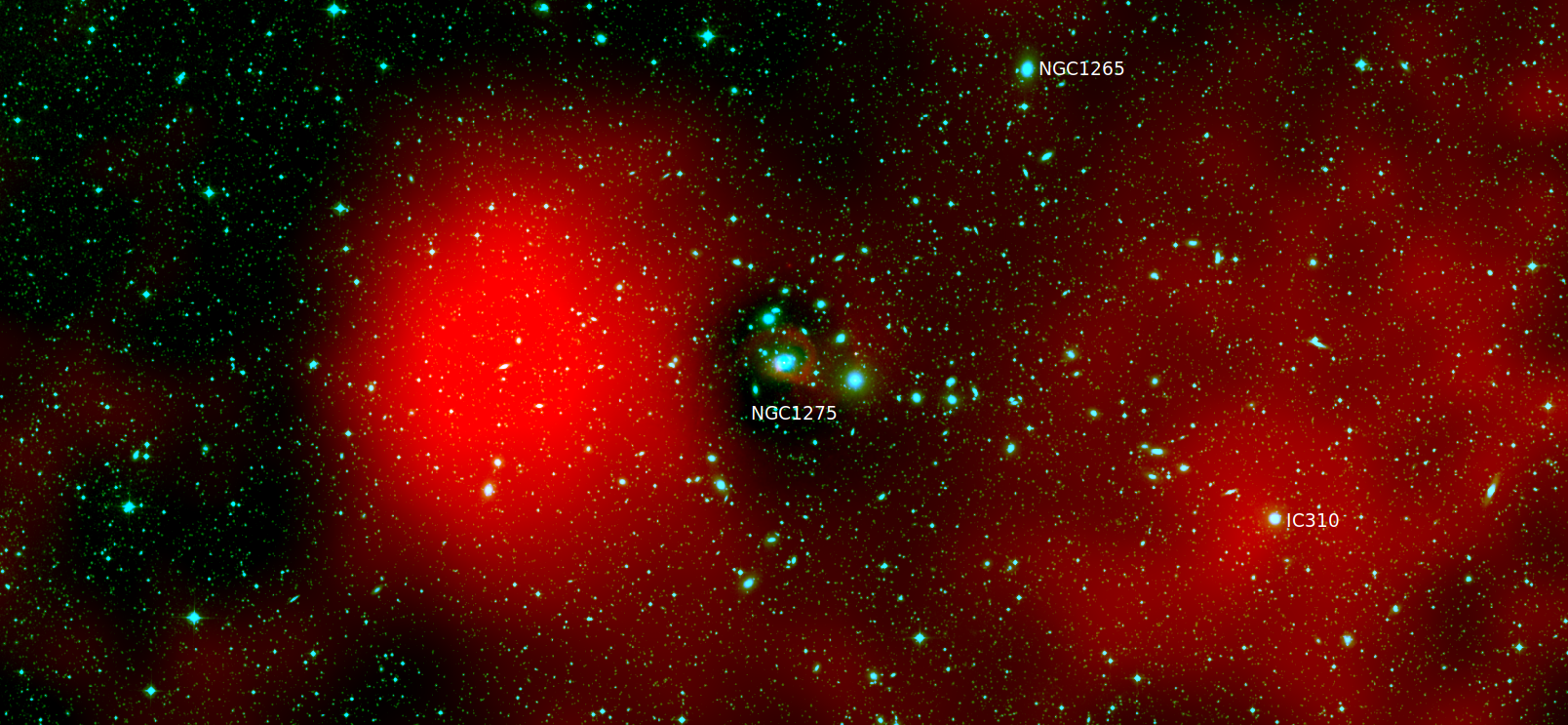}
\caption{Asymmetric structures in a portion of the X-ray image (smoothed version of Fig.~\ref{f:divc}) in red and the DSS2-R image in light blue. The DSS image has been lightly processed to emphasize diffuse sources (galaxies) relative to numerous compact sources (e.g., stars). The chain of galaxies that begins with NGC1275 extends up to IC310, which appears to be at the peak of extended X-ray emission to the West of the cluster center. Further to the West (outside of the image), there is a cluster  AWM7/2A0251+413 at a redshift similar to the Perseus cluster.}
\label{f:xopt}
\end{figure*}

\begin{figure*}
\sidecaption
\includegraphics[angle=0,clip,trim=1.5cm 5.2cm 2cm 4cm,width=12cm]{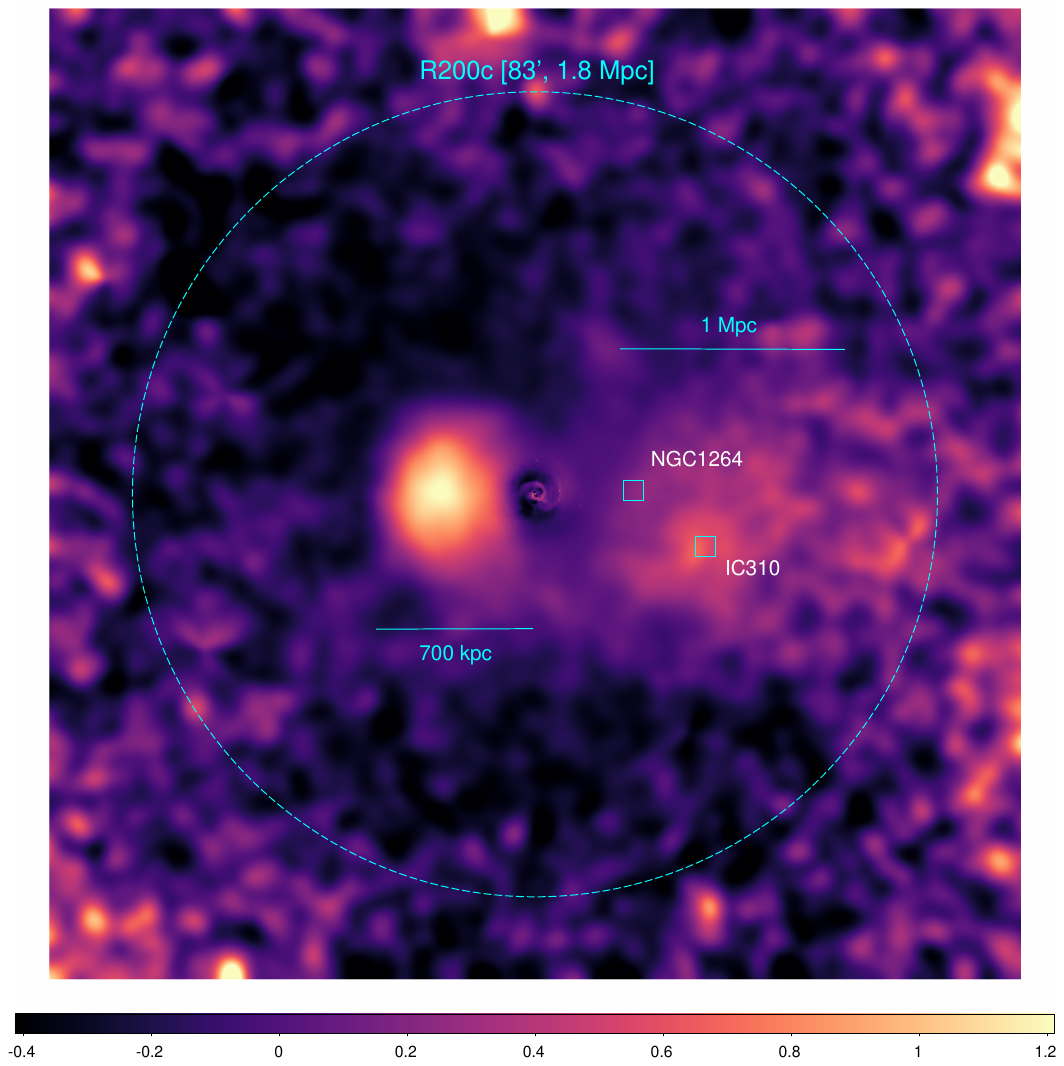}
\caption{Similar to Fig.~\ref{f:divc}, except the eROSITA image has been smoothed with a Gaussian ($\sigma=200''$) after masking the most prominent compact sources. The image size is $200'\times 200'$, i.e., $\sim 4.3\,{\rm Mpc}$ across. The excess emission has a linear size of 1-2~Mpc and extends at least to the Perseus virial radius. Two boxes show the positions of galaxies, IC310 and NGC1264. The latter was associated with a mass concentration based on the weak lensing analysis \cite[see][]{2025NatAs...9..925H}. 
} 
\label{f:divc_200}
\end{figure*}

\begin{figure*}
%\centering
\sidecaption
\includegraphics[angle=0,clip,width=12cm]{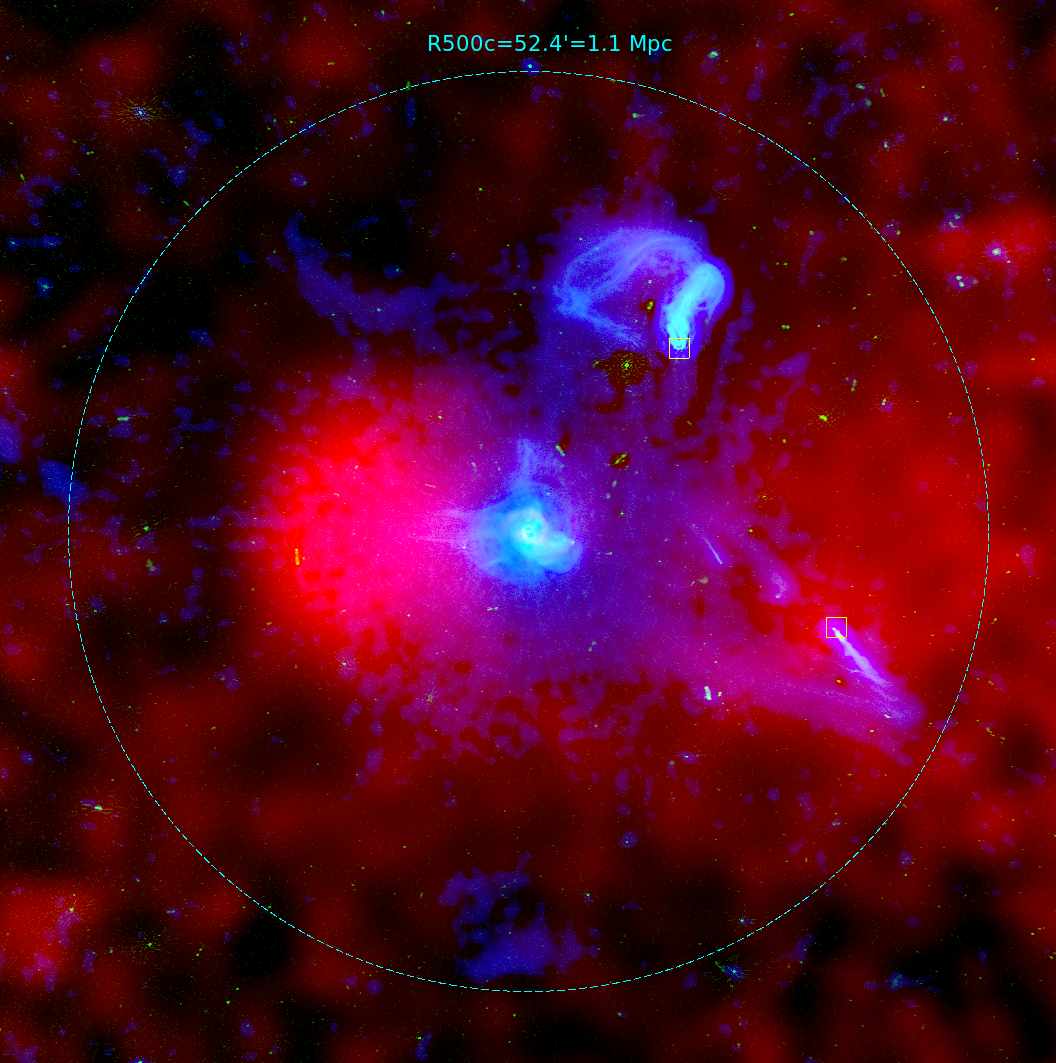}
\caption{2x2 degrees image showing a combination of the X-ray residuals (red, see Fig.~\ref{f:divc}) and radio emission by LOFAR from \cite{2024A&A...692A..12V}. The radio image (blue-cyan colors) is itself composed from the $7''$ (blue) and $80''$ (green) resolution data. NGC1275 mini-halo (and NGC1272) dominates the center. The two most prominent off-center (tailed) galaxies are IC310 and NGC1265 \citep[see][for recent high-quality radio images of the Perseus cluster]{2020MNRAS.499.5791G,2024A&A...692A..12V} }  
\label{f:xradio}
\end{figure*}

\section{Discussion}

\subsection{Merger geometry and timing}
The dynamic state of the Perseus cluster (outside the very core) is a subject of many studies over several decades, 
\citep[see, e.g.,][among others]{1981ApJ...248...55B,1983AJ.....88..697K,1994A&A...284..801A,2003ApJ...590..225C,2011MNRAS.418.2154F,2012ApJ...757..182S,2024ApJ...974..234B,2020A&A...633A..42S,2024ApJS..272...22K,2025NatAs...9..925H} with an overall conclusion that the cluster is neither fully relaxed nor in a state of a major merger. 

The residual X-ray images (Figs.~\ref{f:divc} and \ref{f:divc_200}) show that the excess X-ray emission is well centered at IC310. IC310 is an S0 galaxy that hosts an AGN (BL Lac) that is visible from the radio to TeV bands \citep[e.g.][]{2019MNRAS.485.3277G}. Due to the X-ray bright AGN, the constraints on the diffuse X-ray emission in the vicinity of IC310 are marginal, although the presence of the excess emission (or substructure) on somewhat larger scales has been found \citep[see][]{1992A&A...256L..11S,2005PASJ...57..743S,2010MNRAS.404..180D,2012ApJ...757..182S}. eROSITA data support the interpretation that IC310 is the center of a subcluster merging with the Perseus cluster. Further support for this statement comes from a chain of bright galaxies that apparently "connect" NGC1275 and IC310 (see Fig.~\ref{f:vel_map}).  Yet another piece of evidence comes from radio data, which we discuss in Sect.~\ref{s:2gal}. 

IC310 is located $\sim 800\,{\rm kpc}$ (in projection) to the West of NGC1275. This galaxy has a relatively low line-of-sight velocity compared to the cluster $\sim 600\,{\rm km\,s^{-1}}$,  which is about half of the circular speed in the main cluster. 
In the merger scenario, the perturbing subcluster spends, on average, more time near the apocenter than near the pericenter. If this is the case, the low velocity is expected, and no strong constraints on the true 3D position of IC310 can be placed\footnote{ 
A similar merger configuration is suggested for the Coma cluster, where the NGC4839 group (merging subcluster) has a small line-of-sight velocity relative to the main cluster \citep[e.g.][]{2019MNRAS.485.2922L}.}. At the same time, the lack of obvious gradients in redshifts of galaxies along the main filament (Fig.~\ref{f:vel_map}) suggests that the merger direction is not far from the sky plane, in which case the 3D velocity might be higher.
While IC310 itself does not directly constrain the "age" of the merger, one can hope to use a substructure in the ICM distribution that retains some memory of past perturbations. Those include merger shocks and sloshing of the core gas. 

The association of the prominent asymmetric perturbations (beyond the region strongly dominated by AGN Feedback) in the Perseus cluster \citep[see, e.g.,][]{1992A&A...256L..11S,2012ApJ...757..182S}  with gas sloshing and the Brunt-Vaisala frequency became clear \citep[see Fig.~12 in][]{2003ApJ...590..225C} once the first reliable ICM temperature maps with sufficient angular resolution became available. These maps helped to differentiate between shocks, AGN-inflated bubbles, and contact discontinuities. Using numerical simulations, \cite{2024MNRAS.529..563R} calibrated the effective "propagation velocity" of the contact discontinuity pattern for a Perseus-like cluster. This relation predicts that $R_{\rm CD}  \sim 0.15 c_s t$, where $c_s$ is the ICM sound speed and $t$ is the time since the initial perturbation \cite[see][for discussion of limitations of this expression]{2024MNRAS.529..563R}. Assuming that this relation applies to the prominent excess to the East of the center (see Fig.~\ref{f:divc}), we can estimate the age of this structure. 
Adopting $c_s=1200\,{\rm km\,s^{-1}}$ and the distance from the Perseus core (= NGC1275) of 700~kpc, one gets $t\sim 3.8\,{\rm Gyr}$\footnote{This is somewhat shorter than $t\sim 5.5\,{\rm Gyr}$ suggested by \cite{2025NatAs...9..925H}, where the weak-lensing-based mass concentration associated with NGC1264 (a galaxy NGC1262 some 400 kpc to the West of NGC1275) was suggested as a center of the local mass concentration.}. 

A weak shock propagating with velocity $\sim c_s$ during this time would be at the distance of $\sim 5\,{\rm Mpc}$, well beyond the region studied here. The characteristic time for the subcluster motion (= IC310 in our model) is probably a multiple of the viral time $t=R_{200c}/V_{200c}\sim 1.4\,{\rm Gyr}$. Therefore, it is plausible that the merger shock associated with the pericenter passage that initiated the most prominent contact discontinuity to the East of NGC1275 is long gone, while the IC310 group could be currently at its second (or third) apocenter \citep[see, also, simulations of][]{2024ApJ...974..234B}. Further support for the "near apocenter" scenario is discussed in Sect.~\ref{s:2gal}.

The merger scenario with the excess mass westwards of NGC1275 is also supported by the recent weak lensing data \citep{2025NatAs...9..925H}, albeit the suggested centroid (near the galaxy NGC1264) is a factor of $\sim 2$ closer to the cluster center than IC310. The most recent Euclid data \citep{2025A&A...697A..13K,2025A&A...697A..12M} also show the excess of the intracluster light and dwarf galaxy to the west of the core.

\subsection{Substructure near the virial radius}
At larger scales, comparable to the virial radius, some structures have been reported based on the X-ray data of XMM-Newton and Suzaku. In particular, \citep{2022ApJ...929...37W} suggested two X-ray surface brightness edges at 1.2 and 1.7~Mpc to the west of the core, which were interpreted as cold fronts. \cite{2020MNRAS.498L.130Z} argued that these cold fronts (= contact discontinuities) at very large distances from the center might not be the sloshing structures but rather the result of the collision between the accretion shock and a 'runaway' merger shock \citep[see also][]{2010MNRAS.408..199B}. The analysis of the additional Suzaku data also revealed a discontinuity near $R_{200c}$ (to the North-West from the core), but from the projected temperature profile, it was concluded that this structure is an $M\sim1.9$ shock \citep{2021A&A...652A.147Z}. The eROSITA broad-band data confirm the presence of the substructure in this direction, although the analysis of the spectral characteristics will be reported elsewhere.    

The smoothed eROSITA+Chandra+XMM-Newton residual map\footnote{Only the eROSITA data are smoothed in this combined image to retain the good angular resolution in the core.} is shown in Fig.~\ref{f:divc_200}. This image does show the excess emission westwards to the core extending to the virial radius. Some possible structures near $R_V$ are also seen in the NW direction.  We defer the interpretation of these features for future studies. Here, we only argue that in the minimalists' scenario, these features are plausibly associated with the same merger event. We also note that there is marginal evidence for other large-scale features in the residual X-ray image (Fig.~\ref{f:divc}), e.g., an arc-like feature running almost vertically through NGC1265. However, more data are needed to assess its reality.

\subsection{Two prominent radio galaxies}
\label{s:2gal}
Apart from the central Perseus galaxy, famous NGC1275, the Perseus cluster hosts many prominent radio galaxies \citep[see, e.g.,][for recent VLA nd LOFAR maps]{2020MNRAS.499.5791G,2024A&A...692A..12V}. Here, we focus on two of them - NGC1265 and IC310.  The LOFAR radio image (a combination of $7''$ and $80''$ resolution maps from \citealt{2024A&A...692A..12V}) superposed on the X-ray excess emission image is shown in Fig.~\ref{f:xradio}. 

\subsubsection{IC310} 
As mentioned above, IC310 appears to be at the center of the subcluster/group that merges with the main cluster. In terms of IC310 offset position ($\approx37'=0.44\,R_{200c}$ from the core) and low line-of-sight velocity ($\Delta \varv\simeq 600\,{\rm km\,s^{-1}}=0.48 V_{200c}$ relative to NGC1275)\footnote{Quoted relative line-of-sight velocities are calculated as $\Delta \varv =(\varv_{\rm Gal.}-\varv_{\rm NGC1275})/(1+z_{\rm cl})$.}, this galaxy is reminiscent of the NGC4839 group in the Coma cluster. The low velocity might be related to the proximity of the group to the apocenter, hence low 3D and line-of-sight velocities. While IC310 is originally classified as a narrow-tail galaxy, which would be more consistent with a high relative velocity of the galaxy relative to ICM \citep{1979Natur.279..770B}, it might be a special case of a radio galaxy seen nearly edge-on \cite{1998A&A...331..901S,2020MNRAS.499.5791G}. This interpretation is supported by the blazar-like properties of the nucleus \citep[e.g.][]{2012A&A...538L...1K,2014Sci...346.1080A}, with a pc-scale jet making a relatively small angle to the line of sight\footnote{Another interesting detail discussed by \cite{2012A&A...538L...1K} is that the orientation of the pc-scale jet is similar to that of the kpc-long tail.}. The most recent LOFAR data revealed a much more extended radio tail of IC310 \citep{2024A&A...692A..12V}. The tail makes a sharp $135^\circ$ turn towards the Perseus center (see Fig.~\ref{f:xradio}). This morphology could be explained by a merger scenario, where IC310 has already gone through the pericenter and the apocenter and is now moving back towards the core. In this case, the "reversal" in the far tail direction is explained by the so-called slingshot effect \citep{2019MNRAS.485.2922L,2019ApJ...874..112S,2019MNRAS.488.5259Z}, when the ICM shifted backwards by the ram pressure, swings to the opposite side when the motion of the halo slows down, and then reverses near the apocenter. This configuration leads to a loop-like geometry of the tail that we dubbed "radio-Uroboros". 

Overall, the optical, radio, and X-ray data support the IC310-related merger scenario, happening along the direction of the major East-West filament crossing the Perseus cluster.

\subsubsection{NGC1265} 
Unlike IC310, NGC1265 has a very high velocity relative to NGC1275 ($\Delta \varv\simeq2450\,{\rm km\,s^{-1}}=1.94 V_{200c}$) and it lacks obvious large-scale X-ray emission around the X-ray bright core. While there are even higher velocity galaxies in the same region (plausibly associated with another large-scale filament, see Fig.~\ref{f:vel_map}), the radio tail suggests interaction with the Perseus ICM. The tail identified in \cite{1968MNRAS.138....1R,1975A&A....38..381M}, led to the classification of NGC1265 as a Narrow-angle-tail galaxy, although at higher resolution, the core rather looks like a wide-angle-tail galaxy \citep[see a discussion of tags and boxes in][]{2021Galax...9...85R}. The extent of its long and bent tail (projected length $\sim 800\,{\rm kpc}$) was revealed with WSRT observations \citep{1998A&A...331..901S} and more recently imaged in great detail \citep{2020MNRAS.499.5791G,2024A&A...692A..12V}.

Remarkably, despite being very different from IC310, the radio tail of NGC1265 also qualifies for the "radio-Uroboros" tag (see Fig.~\ref{f:xradio}). The statistics of radio tails' direction suggest that for a bare galaxy, especially moving with a very high velocity, the gas feeding the central AGN can be stripped during the first infall \citep[e.g.,][]{2020MNRAS.495..554R,2022ApJ...934..170L}, NGC1265 appears to be a newcomer to the Perseus cluster. Yet, it has a very extended tail (probably a factor of a few longer than its apparent projected size). In \cite{2011ApJ...730...22P}, the curious properties of the NGC1265 tail are explained as due to the galaxy crossing an accretion shock near the virial radius, with the galaxy located at the near side of the cluster.  While this is a plausible scenario, we note that the ICM in the Perseus can be significantly perturbed by the ongoing merger, and for a galaxy on a nearly straight trajectory, dictated by its high velocity, the ICM motions can perturb the radio tail left by NGC1265. We, therefore, consider below a simple model with NGC1265 at the far side (along the line of sight) of the main cluster. This model broadly corresponds to scenarios c) and d)  in \cite{1998A&A...331..901S}.

\subsubsection{Test particle (NGC1265) in a static potential}
\label{s:orbit}

\begin{figure}
\centering
\includegraphics[angle=0,clip,trim=0.5cm 5.5cm 1cm 2.5cm,clip,width=0.95\columnwidth]{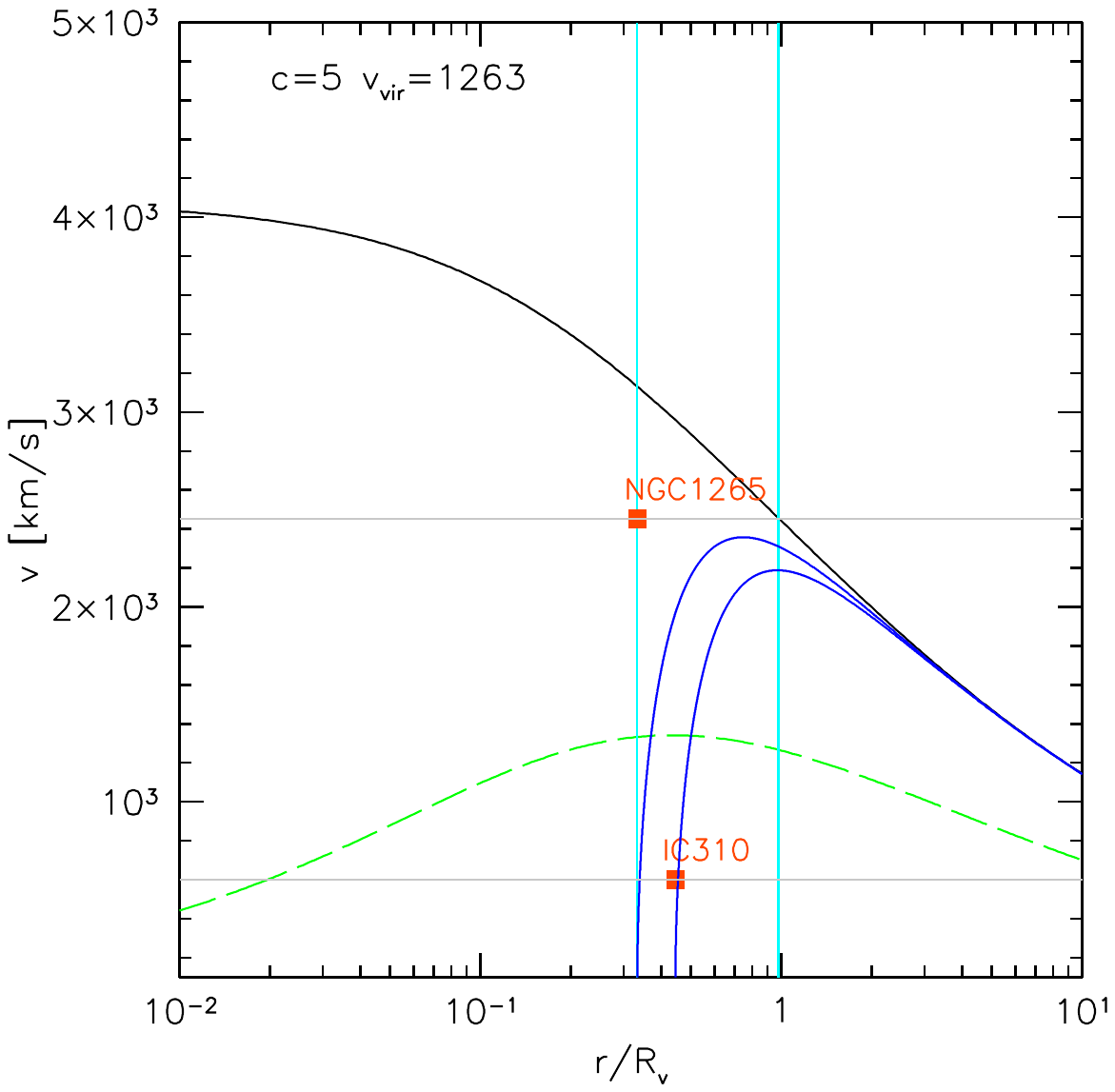}
\caption{Velocity of a marginally bound point mass (zero total energy, i.e., nearly parabolic orbit) in a static NWF halo with the concentration parameter $c=5$ and virial velocity $V_{\rm vir}=1263 \,{\rm km\,s^{-1}}$ as a function of radius (black curve). The red points show the line-of-sight velocities  ($\varv_{\rm los}$) and projected distances $r_p$ of the radio galaxies IC310 and NGC1265. The green line shows the circular velocity at a given radius. The pair of blue curves shows the expected l.o.s. velocity of a point mass on a radial trajectory as a function of 3D radial distance from the center for the known projected radii of these two galaxies.  For IC310, its small line-of-sight velocity is fully consistent with being almost in the picture plane, i.e., the projected distance can be almost equal to the true 3D distance from the cluster core. For NGC1265, the two vertical cyan lines show the allowed range of 3D distance, set by the requirement that the observed line-of-sight velocity is smaller than the 3D velocity of a marginally bound point mass.  
}
\label{f:vlos}
\end{figure}

\begin{figure}
\centering
\includegraphics[angle=0,clip,trim=0.5cm 5.5cm 1cm 2.5cm,width=0.99\columnwidth]{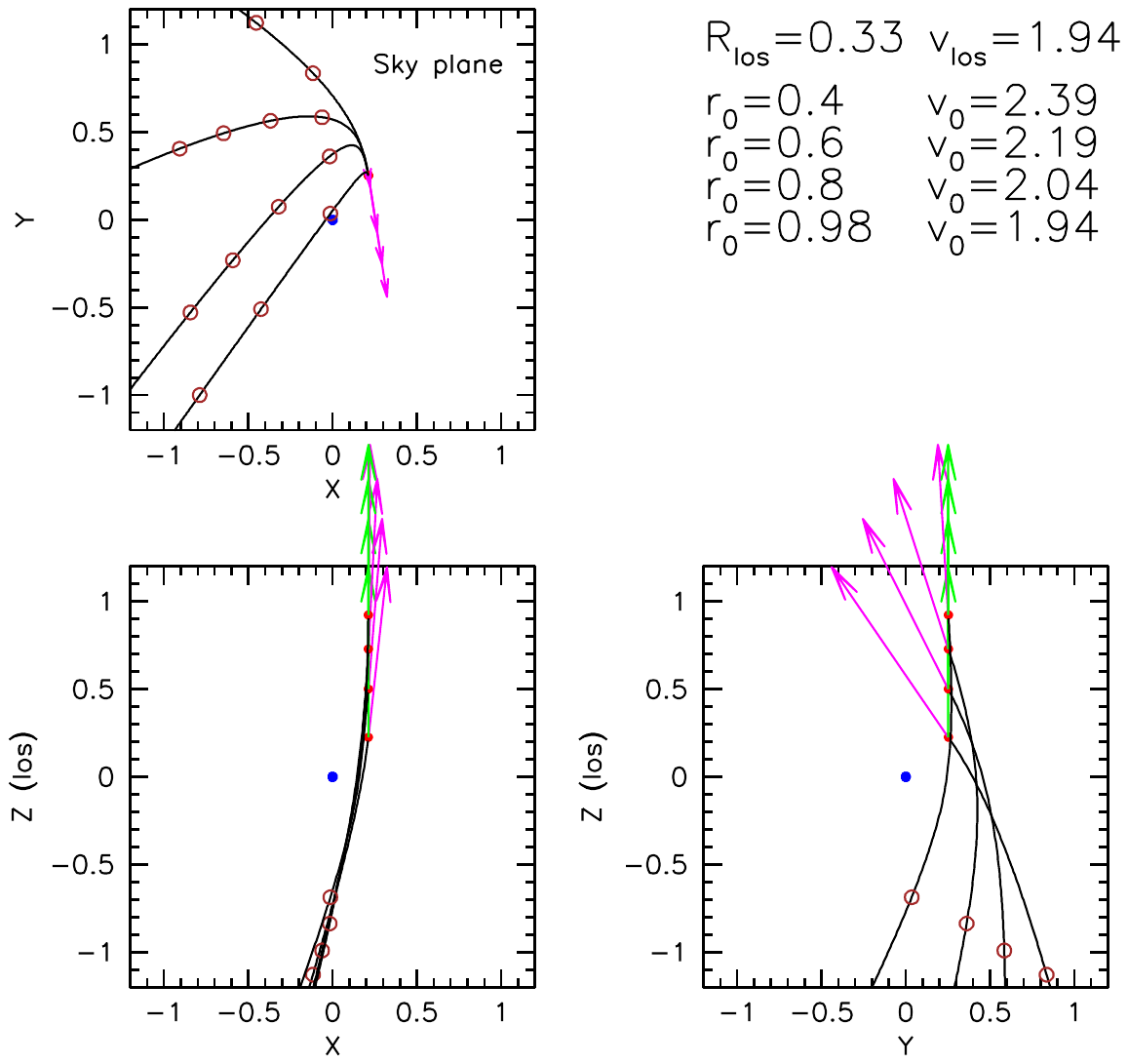}
\caption{Sample trajectories of a marginally bound point mass (NGC1265) in a static NFW potential. Three projections are shown. Here, X and Y are the coordinates in the sky plane, and Z is along the line of sight. 
The distances and velocities are in A426 virial units,   1.79~Mpc and $1263\,{ \rm km\,s^{-1}}$, respectively. 
The blue and red dots mark the positions of the cluster center and the galaxy (current location), respectively. Brown circles mark the galaxy position $n\times$Gyr ago ($n=1,2,...$). The current projected distance and the line-of-sight velocities are fixed at the observed values. Also fixed is the current direction of the projected velocity, which is guessed from the morphology of the radio emission near the core. Therefore, all trajectories shown have the same sky-plane velocity direction but different 3D distances from the core ($r_0$) and 3D velocities ($\varv_0$). For the largest 3D distance shown, $r_0=0.98$, the galaxy is currently moving almost along the lines of sight ($\varv_0\simeq \varv_{\rm l.o.s.}$), and its trajectory goes through the very core (in projection).
No attempt to further fine-tune the orbital parameters was made. If the observed radio tail global shift (to the East of the galaxy) reflects the overall geometry of the orbit, the location of NGC1265, farther away from us than NGC1275, is more promising. The green arrows show the line-of-sight velocities, while the magenta arrows show the projections of the full 3D velocity on the corresponding planes.} 
\label{f:orbit_ngc1265}
\end{figure}

As the starting point, we consider a test particle moving in a static potential well having an NFW mass distribution \citep{1997ApJ...490..493N}.  We further assume a zero total energy for this particle, i.e, $\varv^2/2+\phi(r)$. 
We chose a coordinate system with $X$ and $Y$ in the sky plane, and $Z$ along the line-of-sight. We aim to reproduce the line-of-sight velocity and the projected distance of NGC1265. The requirement to have zero total energy implies that the 3D distance of the particle from the cluster center is less than $r_{\rm max}=0.98R_V$, while the lower limit is set by the observed projected distance from the cluster center, i.e., $r_{\rm min}=R_p$ (see Fig.~\ref{f:vlos}). If the marginally bound particle is close to $r_{\rm max}$, it has to move exactly along the line of sight. In the opposite case, the angle to the line of sight is set by the relation between the local 3D velocity (set by the 3D radius) and the $\varv_{\rm los}$. For the limiting case of $r\approx r_{\rm min}=R_p$, this angle is about $38$ degrees. For any realistic 3D position (further away than $R_p$), this angle will be smaller. If the overall shift of the tail eastwards of the current position of NGC1265 reflects the large-scale trajectory of the galaxy, it should be crossing the main cluster on a slightly curved trajectory. Given the high velocity of NGC1265, the curved trajectory requires the galaxy to cross the core of the cluster. A sample of such trajectories is shown in Fig.~\ref{f:orbit_ngc1265}. They all correspond to the same (current) projected distance, line-of-sight velocity, and the velocity direction in the sky plane, and differ in the magnitude of the 3D velocity. This is, of course, a small subsample of possible trajectories (given direct observational constraints and our assumption that the galaxy is currently on the far side of the cluster).  In this model, we assume that gas motions in the region between NGC1275 and IC310 further disturb the tail trajectory to give it the observed shape. No attempt was made to get constraints on the detailed properties of the tail, given that the density and pressure of the ICM change significantly along the galaxy trajectory, and the power of the NGC1265 jets can vary too. 

The above model does not consider constraints coming from the uncertainties associated with the Faraday Rotation measurement in the direction of NGC1265 \citep[see, e.g.,][]{2005A&A...441..931D,2011A&A...526A...9B}, that might predominantly come from the Milky Way rather than from the cluster \citep[see][for a scenario with NGC1265 on the near side of the cluster]{2011ApJ...730...22P}.

Another limitation comes from the assumption of a static potential of the main cluster. In reality, peculiar initial velocities and time variations of the potential can play a role. For a merging cluster such as Perseus, the latter effect might be important \citep[e.g.,][]{2007MNRAS.379.1475S,2014JCAP...06..057C}. In this case, the trajectory of the galaxy might be more complicated. We do not discuss this scenario further.

Overall, we see several attractive features of the model described above, but we can not exclude other scenarios.

\subsubsection{Line-of-sight velocities in simulations}
\label{s:magneticum}
While the NGC1265 trajectories discussed in Sect.~\ref{s:orbit} were derived assuming zero binding energy of the galaxy in a static NFW potential, one can pose a question about how often we expect to find a galaxy given its projected distance and the line-of-sight velocity (with both quantities expressed in terms of virial properties of the main halo). Since the answer depends on the mass accretion rate of the main halo and on the importance of other effects such as dynamic friction (for massive subhalos), it can be best addressed by looking at the results of numerical simulations. Here, we use a publicly available catalog of halos in \textit{Magneticum}\footnote{www.magneticum.org} simulations set \citep{2014MNRAS.442.2304H,2016MNRAS.463.1797D,2025arXiv250401061D}, see also \citealt{2023MNRAS.521.3981A}.

These simulations were performed with the TreePM/SPH code \texttt{GADGET-3} \cite[][]{2005MNRAS.364.1105S, 2016MNRAS.455.2110B, 2023MNRAS.526..616G}.
They take into account many complex non-gravitational physical processes (cooling, merging of galaxies and galaxy clusters, shock waves, and detailed galaxy formation physics, including the treatment of black holes) which determine the evolution of large-scale systems and affect their observational properties. More details of the physics provided in the simulation can be found in Section 2 of \cite{2022A&A...661A..17B} and references therein. We take advantage of the haloes catalogue extracted from \textit{Box2/hr} simulation box at its terminal redshift. Its comoving volume is equal to $(352 \text{Mpc}/h)^3$ ($h=H_0/{\rm (100~km s^{-1} Mpc^{-1})}$ is the Hubble constant), containing $2\cdot1584^3$ mass resolution elements (particles). The masses of dark matter and gas particles are equal to $m_{\rm DM} = 6.9\cdot10^8 M_{\odot}/h$ and $m_{\rm gas} = 1.4\cdot10^8 M_{\odot}/h$, respectively, and the following cosmological parameters are adopted: the total matter density $\Omega_{\rm M}=0.272$ ($16.8\%$ baryons), the cosmological constant $\Omega_{\Lambda}=0.728$, the Hubble constant $H_0=70.4$ km/s/Mpc (i.e. $h=0.704$), the index of the primordial power spectrum $n=0.963$, the overall normalization of the power spectrum $\sigma_8=0.809$ \citep{2011ApJS..192...18K}.

\begin{figure}
\centering
\includegraphics[angle=0,clip,trim=0.5cm 5.0cm 1cm 2.5cm,clip,width=0.95\columnwidth]{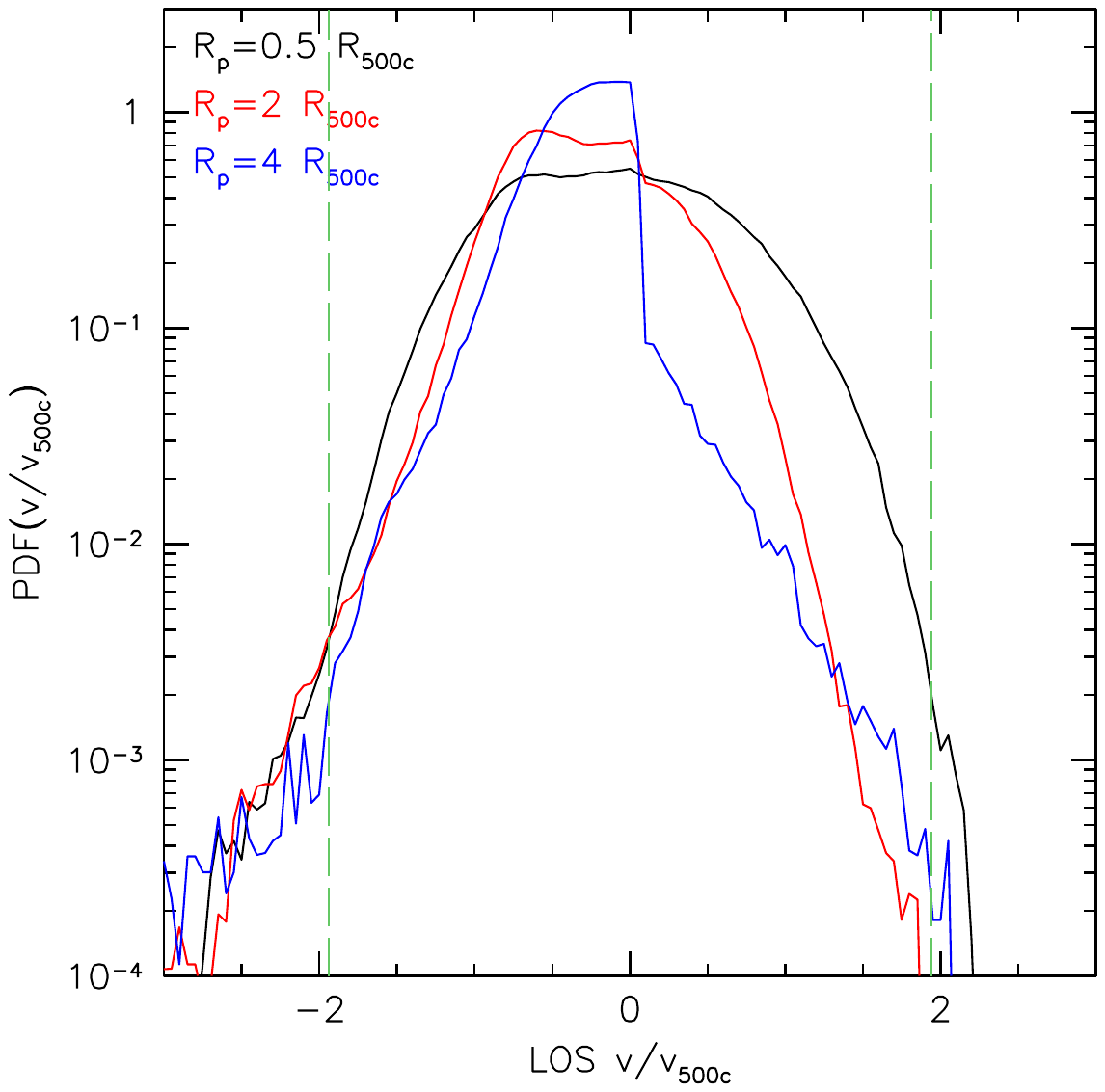}
\caption{Probability of having line-of-sight velocity $\sim 2500\, {\rm km\,s^{-1}}$ of a galaxy in the vicinity of a massive cluster for different projected distances $R_p$ (based on Magneticum simulations, see Sect.~\ref{s:magneticum}). The velocity is normalized by the circular speed of the cluster $V_{500c}\approx V_{200c}$. The sign of the velocity is set by the sign of the radial velocity component (negative - infalling). The observed position of NGC1265 corresponds to the black curve ($R_p\sim 0.5R_{500c}$). At this projected distance, the ${\rm PDF(V)}$ is rather similar for galaxies moving towards the cluster center or away from it. The green/dashed vertical lines show the observed NGC1265 velocity in units of $V_{500c}$. The probability of finding such a high velocity is  $\sim 0.1$\% - low but not prohibitively low. The Hubble expansion is not added to galaxies' velocities.}
\label{f:pdfv_rp}
\end{figure}

\begin{figure}
\centering
\includegraphics[angle=0,clip,trim=0.5cm 5cm 1cm 2.5cm,clip,width=0.95\columnwidth]{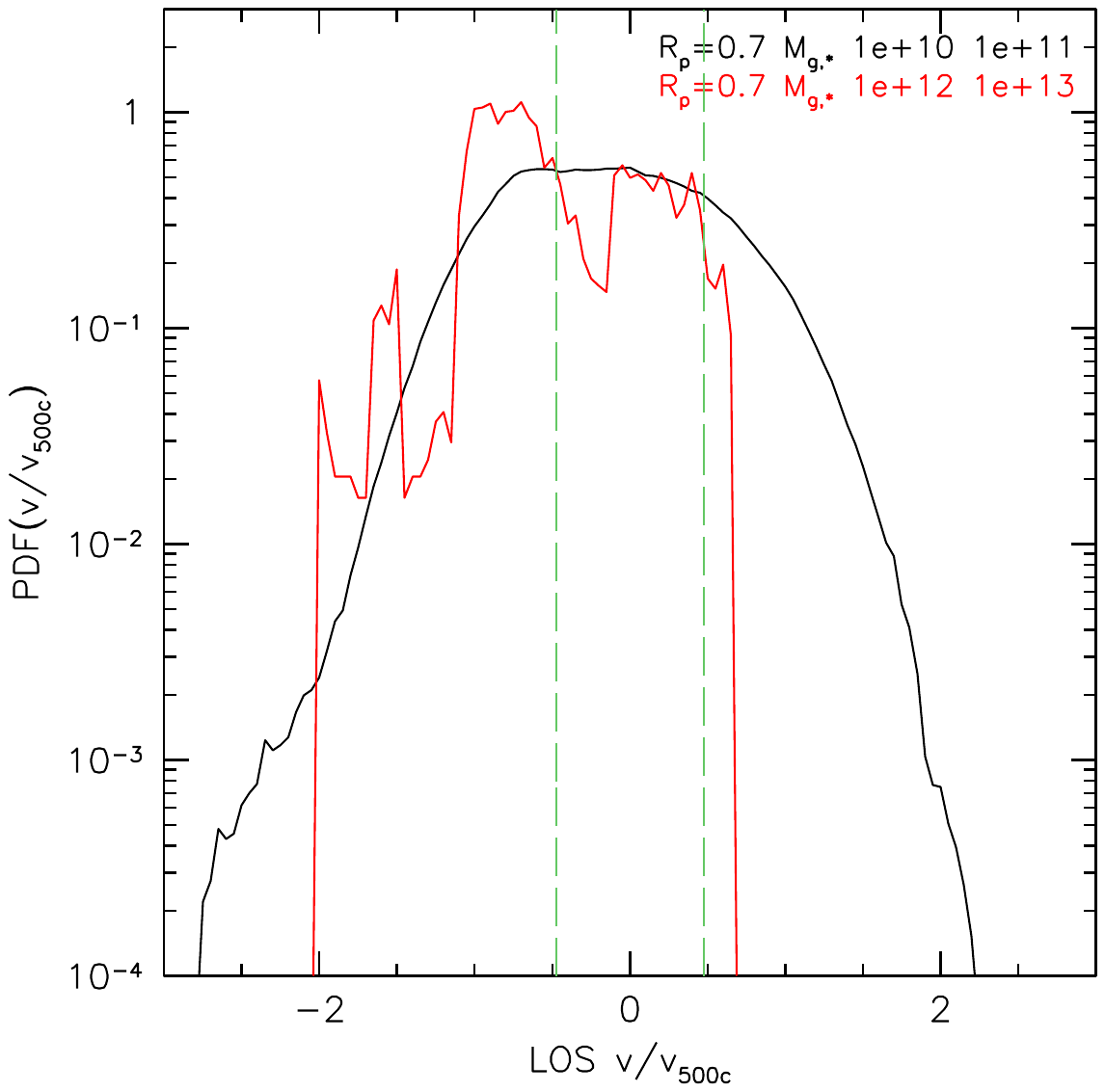}
\caption{IC310 case - massive group at projected distance $R_p=0.7 R_{500c}$. The plot is similar to Fig.~\ref{f:pdfv_rp} but now the red curve shows ${\rm PDF}(V)$ for massive subhalos with stellar mass in the range $10^{12}-10^{13}\,M_\odot$. This distribution is now very asymmetric between infalling and outgoing halos, presumably due to dynamic friction. The IC310 has a small line-of-sight velocity, consistent with an object not far from the pericenter. The Hubble expansion is not added to galaxies' velocities.  
}
\label{f:pdfv_mass}
\end{figure}

For each halo more massive than $7\times 10^{13}\,M_\odot$, we identified all subhalos. 
The distances (from the halo to subhalos) and velocities have been rescaled to $R_{500c}$ and $V_{500c}$ of the main halo\footnote{We used the "500c"-based quantities (rather than "200c") since they are provided in the catalogs. $V_{500c}$ is close to $V_{200c}$ anyway, while the relation between $R_{500c}$ and $R_{200c}$ changes from 0.63 to 0.66 for concentration parameter $c$ between $3$ and $5$}. This information was used to construct the probability density function of the line-of-sight velocity as a function of projected distance. Namely, each subhalo, located at a distance of $r$ from the main halo, was accounted for when considering all possible viewing angles corresponding to projected distances from $R=0$ to $R=r$ and the corresponding solid angles. We also set the sign of the line-of-sight velocity based on the sign of the radial component of the velocity. Negative values correspond to subhalos currently moving towards the main halo\footnote{Such information is not available in observations.}.  The resulting PDF is shown in Fig.~\ref{f:pdfv_rp} for three values of projected distances. As expected, at large projected distances, the PDF is asymmetric due to the "first-infall" region of the phase space, while at smaller distances this difference becomes less pronounced.   The two vertical dashed lines show the line-of-sight velocity of NGC1265, which is $\sim 1.94 V_V$. The projected distance of NGC1265 from the Perseus core is $R\sim 0.5R_{500c}$ (the black curve in Fig.~\ref{f:pdfv_rp}. The probability of finding a galaxy at this projected distance with the line-of-sight velocity larger than $\sim 2 V_{200c}$ is less than 1\% (actually, close to 0.1\%). This makes NGC1265 a very rare object, but not truly exceptional.

We apply the data from the same catalogues to the IC310 case, but this time we fix the projected distance and plot in Fig.\ref{f:pdfv_mass} the velocity PDF separately for low mass galaxies and more massive objects (groups), implicitly assuming that IC310 is indeed the central galaxy of a subcluster and belongs to the "massive" subsample. This separation was based on the stellar mass, because this quantity is a somewhat better-defined (and conserved) quantity than the total mass. The high mass bin effectively contains all massive halos, i.e., clusters themselves, but they do not contribute to the plotted distribution since they have zero offset from the parent cluster.   The projected distance of IC310 is $\sim 0.7R_{500c}$, while its velocity is $\sim 0.5 V_{200c}$ (see two vertical green lines in  Fig.\ref{f:pdfv_mass}). For the low mass sample (the black curve), the distribution is symmetric, and IC310 (if it is an individual galaxy rather than the center of the group) is in the central part of the distribution and has an equal probability of being infalling and outgoing. For the high mass case (the red line), the observed line-of-sight velocity is also consistent with both cases, while a factor of $\sim 1.5$ larger observed velocity would already be in tension with the outgoing scenario. Based on these estimates, we concluded that the infalling and outgoing scenarios can not be excluded for IC310, even if it is a group of galaxies.      

\section{Conclusions}
\label{sec:conclusions}
We combined Chandra, XMM-Newton, and eROSITA data to make a complete image of the Perseus cluster from the very core to the virial radius. This image shows all the spectacular features that a bona fide cluster might have: clear signs of AGN Feedback in the core, Cold Fronts (contact discontinuities) at intermediate scales, and signatures of a merger at the largest scales. 

eROSITA X-ray data suggest that the IC310 galaxy ($\sim800\,{\rm kpc}$ west of NGC1275) is at the center of the subcluster/group that perturbed Perseus some 4~Gyr ago. A clear excess of X-ray emission is seen in a $\sim$Mpc-size region around IC310, implying that the subcluster was able to retain some of its gaseous atmosphere.  

IC310 has a rather small recession velocity $\varv\sim 600\,{\rm km\,s^{-1}}\sim 0.5 V_{\rm 200c}$ relative to the cluster. We tentatively attribute this small velocity to the selection effects - it is more likely to find the infalling object near the apocenter than near the pericenter. This is, of course, a qualitative argument. It is supported by the shape of the IC310 radio tail, which bends sharply in the vicinity of the galaxy. Such a sharp turn could arise when the sign of the radial velocity changes near the apocenter. Alternatively, the low velocity of IC310 could reflect the merger direction in the sky plane, given that only a weak gradient in recession velocities is seen in the major large-scale filament running through Perseus.

At larger scales, the excess X-ray emission can be traced westwards up to the virial radius, with some tentative arc-like features in the NW direction. It is plausible that these structures are associated with the same merger.  Independent of the details, the elongation of the Perseus cluster X-ray emission in the East-West direction coincides well with the orientation of the large-scale filament of the Perseus-Pisces supercluster.

While this paper was under review, two new studies of the Perseus cluster based on the XRISM observations \citep{2025arXiv250904421X,2025arXiv251012782Z} became available. Broadly, these studies are consistent with the scenario discussed here. 

Finally, we discussed briefly the case of the high-velocity radio galaxy NGC1265, known for its curious radio tail, which also shows sharp bends. In our "minimalist" model, this galaxy is moving nearly along the line of sight with the total 3D velocity $\sim 2V_{200c}$ and is currently on the far side of the main cluster. The gas motions induced by the IC310 merger might play a role in shaping its tail, although other explanations are possible, too.

\begin{acknowledgements}
    
This work is partly based on observations with the eROSITA telescope onboard \textit{SRG} space observatory. The \textit{SRG} observatory was built by Roskosmos in the interests of the Russian Academy of Sciences, represented by its Space Research Institute (IKI), in the framework of the Russian Federal Space Program, with the participation of the Deutsches Zentrum für Luft- und Raumfahrt (DLR). The eROSITA X-ray telescope was built by a consortium of German Institutes led by MPE and supported by DLR. The \textit{SRG} spacecraft was designed, built, launched, and operated by the Lavochkin Association and its subcontractors. The science data are downlinked via the Deep Space Network Antennae in Bear Lakes, Ussurijsk, and Baikonur, funded by Roskosmos. 

The development and construction of the eROSITA X-ray instrument was led by MPE, with contributions from the Dr. Karl Remeis Observatory Bamberg $\&$ ECAP (FAU Erlangen-Nuernberg), the University of Hamburg Observatory, the Leibniz Institute for Astrophysics Potsdam (AIP), and the Institute for Astronomy and Astrophysics of the University of Tübingen, with the support of DLR and the Max Planck Society. The Argelander Institute for Astronomy of the University of Bonn and the Ludwig Maximilians Universität München also participated in the science preparation for eROSITA. The eROSITA data were processed using the eSASS/NRTA software system developed by the German eROSITA consortium and analyzed using proprietary data reduction software developed by the Russian eROSITA Consortium.

This publication makes use of data products from the Two Micron All Sky Survey, which is a joint project of the University of Massachusetts and the Infrared Processing and Analysis Center/California Institute of Technology, funded by the National Aeronautics and Space Administration and the National Science Foundation.

IK and KD acknowledge support by the COMPLEX project from the European Research Council (ERC) under the European Union’s Horizon 2020 research and innovation program grant agreement ERC-2019-AdG 882679.

The calculations for the hydrodynamical simulations were carried out at the Leibniz Supercomputer Center (LRZ) under the project pr83li (Magneticum). 
\end{acknowledgements}

\bibliographystyle{aa}
\bibliography{ref}

\begin{appendix}

\section{Chandra, XMM-Newton, and SRG/eROSITA datasets}
\label{a:obsids}
The OBSIDs of publicly available Chandra and XMM-Newton data used in this study are listed below. \\

For Chandra: 
11713, 11714, 11715, 11716, 12025, 12033, 12036, 12037, 13989, 13990, 13991, 13992, 3209, 4289, 4946, 4947, 4948, 4949, 4950, 4951, 4952, 4953, 6139, 6145, 6146. 

Observations were processed with the standard Chandra data reduction (CIAO v. 4.17) and calibration software (CalDB v. 4.12). Data analysis steps are described in detail in  \cite{2009ApJ...692.1033V}  and include high background period filtering, application of the latest calibration corrections to the detected X-ray photons, and determination of the background intensity in each observation. 
In practice, for the imaging analysis of the central region of the Perseus cluster, the details of the background calibration are not important.

\vspace{0.5cm}

For XMM-Newton:
0085110101, 0085590201, 0151560101, 0204720101, 0204720201, 0305690101, 0305690301, 0305690401, 0305720101, 0305720301, 0305780101, 0405410101, 0405410201, 0673020201, 0673020301, 0673020401.

The XMM-Newton data were cleaned for flares and then processed similarly to \cite{2003ApJ...590..225C}, utilizing recent versions of the calibration files. Only the MOS data were used. The steady component of the detector background was subtracted using data from blank field observations. The variable component of the detector background was modeled as a power law as a function of energy $B(E) = AE^{-0.45}$. The
data in the 11-12 keV range were used to calculate the normalization of this variable component. As with the Chandra data, only the central part of the cluster measured by XMM contributes to the combined image, and, therefore, the results are not affected by the treatment of the background.

\vspace{0.5cm}

For eROSITA: data from all four consecutive scans are combined after filtering for the periods of enhanced solar activity, which cause short periods of strongly elevated levels of the instrumental background. For the imaging analysis, the data taken with all seven telescope modules (TMs) are combined. Data reduction, filtering, vignetting-correction, and background subtraction are performed in the same way as was done in the previous studies exploring Galactic diffuse X-ray sources, including those with the surface brightness below the total (sky and detector) background level \citep{2021MNRAS.507..971C,2022MNRAS.509.6068K,
2023MNRAS.521.5536K,2024A&A...689A.278K}.
Specifically, the energy-dependent contribution of the instrumental background is modeled and subtracted based on the calibration data accumulated via observations with the "closed filter wheel" configuration, while corrections for exposure time and vignetting are conducted so that the data are characterized by field-of-view-averaged response matrices. 
Deep observations of the Coma cluster \citep[][]{2021A&A...651A..41C} were instrumental for eROSITA's spectral cross-calibration via ICM temperature measurement in its core, while sets of bright point sources observed during the all-sky survey were used to construct the shape of the off-axis PSF \citep[][]{2023A&A...670A.156C}.

\section{Radially dependent weights}
\label{a:weight}

\begin{figure}
\centering
\includegraphics[angle=0,trim=0cm 5.5cm 1cm 2.5cm,clip,width=0.95\columnwidth]{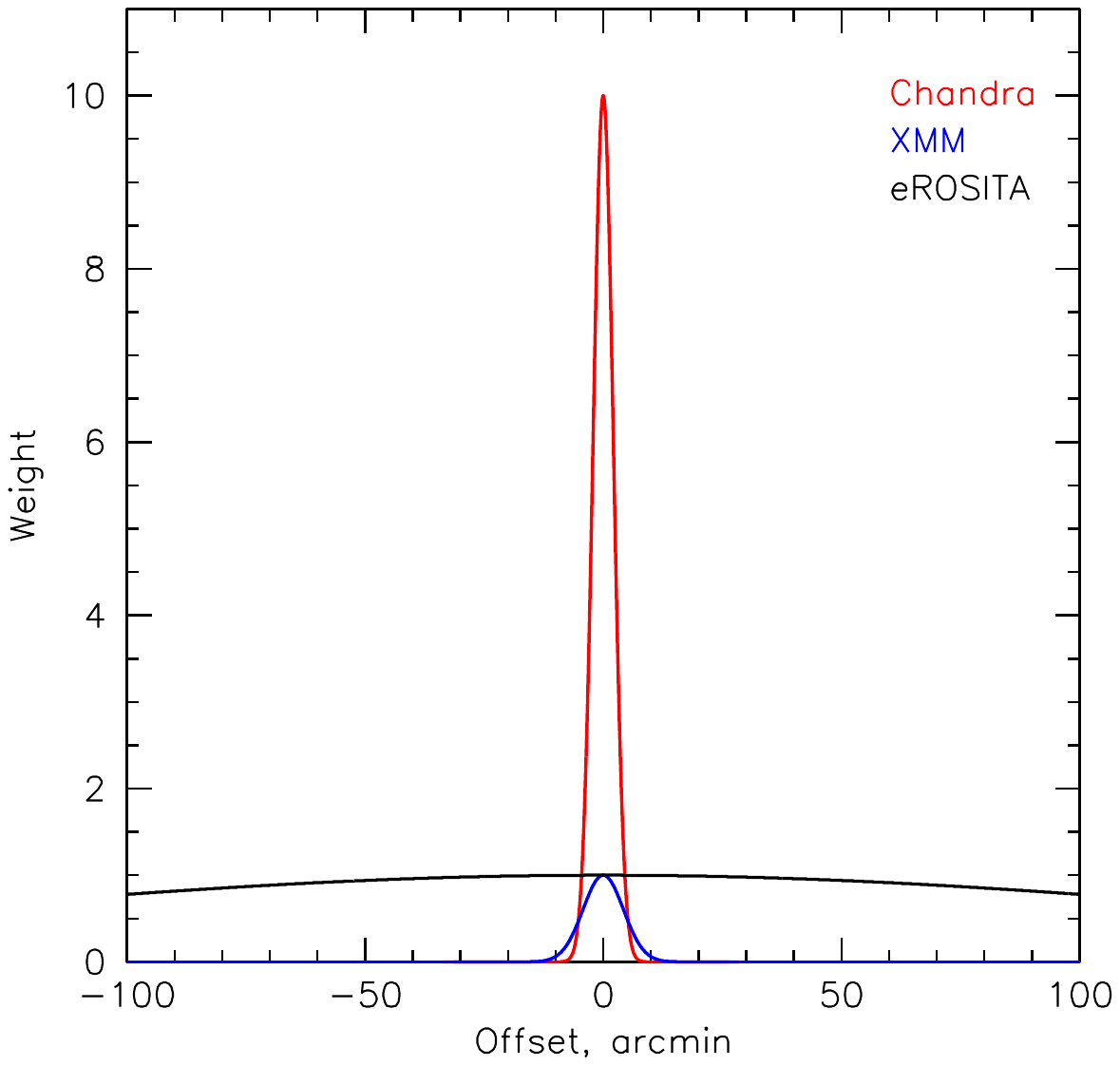}
\caption{Radially-dependent weights used to combine Chandra, XMM-Newton, and eROSITA images. 
The main goals of this weighting scheme are (i) to preserve Chandra's high angular resolution in the core region, where S/N is very high, and (ii) to suppress the noisy edges of the areas covered by Chandra and XMM-Newton. The images were recalibrated to produce the same total flux from the core region covered by all three instruments.
}
\label{f:weights}
 \end{figure}

As described in Sect.~\ref{s:ximage}, eROSITA, Chandra, and XMM-Newton images were rebinned to the same spatial grid (2" pixels) and re-normalized to ensure a common flux calibration. For this re-normalization, a circle ($3'$ radius) centered at NGC1275 was used, and all fluxes were reduced to the eROSITA values. The last step was the coadding of images with radially dependent weights 

\begin{equation}
I(x,y)=\frac{\Sigma_i W_i(r)I_i(x,y)}{\Sigma_i W_i(r)},   
\label{e:weight}
\end{equation}
where $i$ is the image's indices (1,2,3 for Chandra, XMM-Newton, and eROSITA, respectively). The weights are defined as
\begin{equation}
    W_i(r)=A_i e^{-r/R_i},
\end{equation}
where $r$ is the distance from NGC1275 in arcminutes; $R_i=600,3,6$ for eROSITA, Chandra, and XMM, respectively, and $A_i=1,10,1$.

\begin{figure*}
\includegraphics[angle=0,clip,width=2\columnwidth]{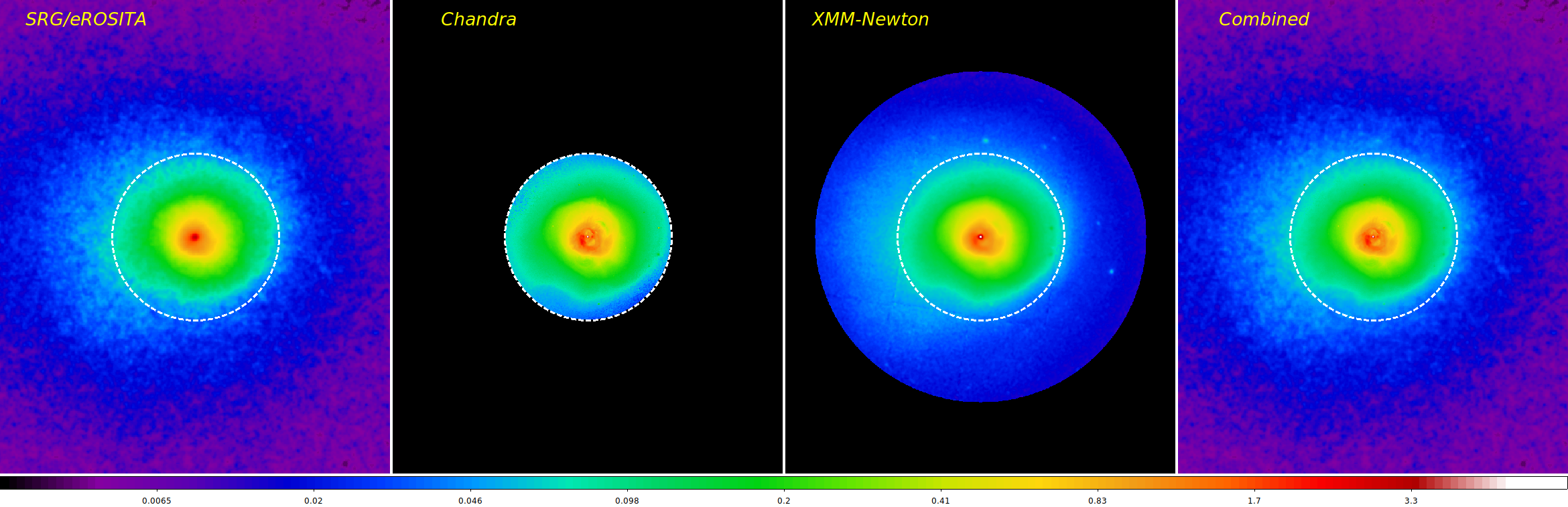}
\caption{X-ray images of the Perseus cluster obtained by eROSITA, Chandra, and XMM-Newton before combining them using the weights described in the text. Only the central part of the image is shown, where the contributions of all three images are important. For Chandra and XMM-Newton, only the parts of the image where the radially-dependent weight $W(r)$ is greater than $\sim 2\times 10^{-2}$ are shown. The dashed circle (shown to facilitate visual comparison) has a radius of $6'$. 
} 
\label{f:ecx_images}
\end{figure*}

The main goal of this weighting scheme is to preserve Chandra's high angular resolution in the core and suppress the noisy contributions at the edges of Chandra and XMM-Newton images. The implemented weights are shown in Fig.~\ref{f:weights}. The individual images are shown in Fig.~\ref{f:ecx_images}.

\section{Radial X-ray surface brightness profile}
\label{a:radial}
\begin{figure}
\centering
\includegraphics[angle=0,trim=1cm 5.5cm 1cm 2.5cm,clip,width=0.95\columnwidth]{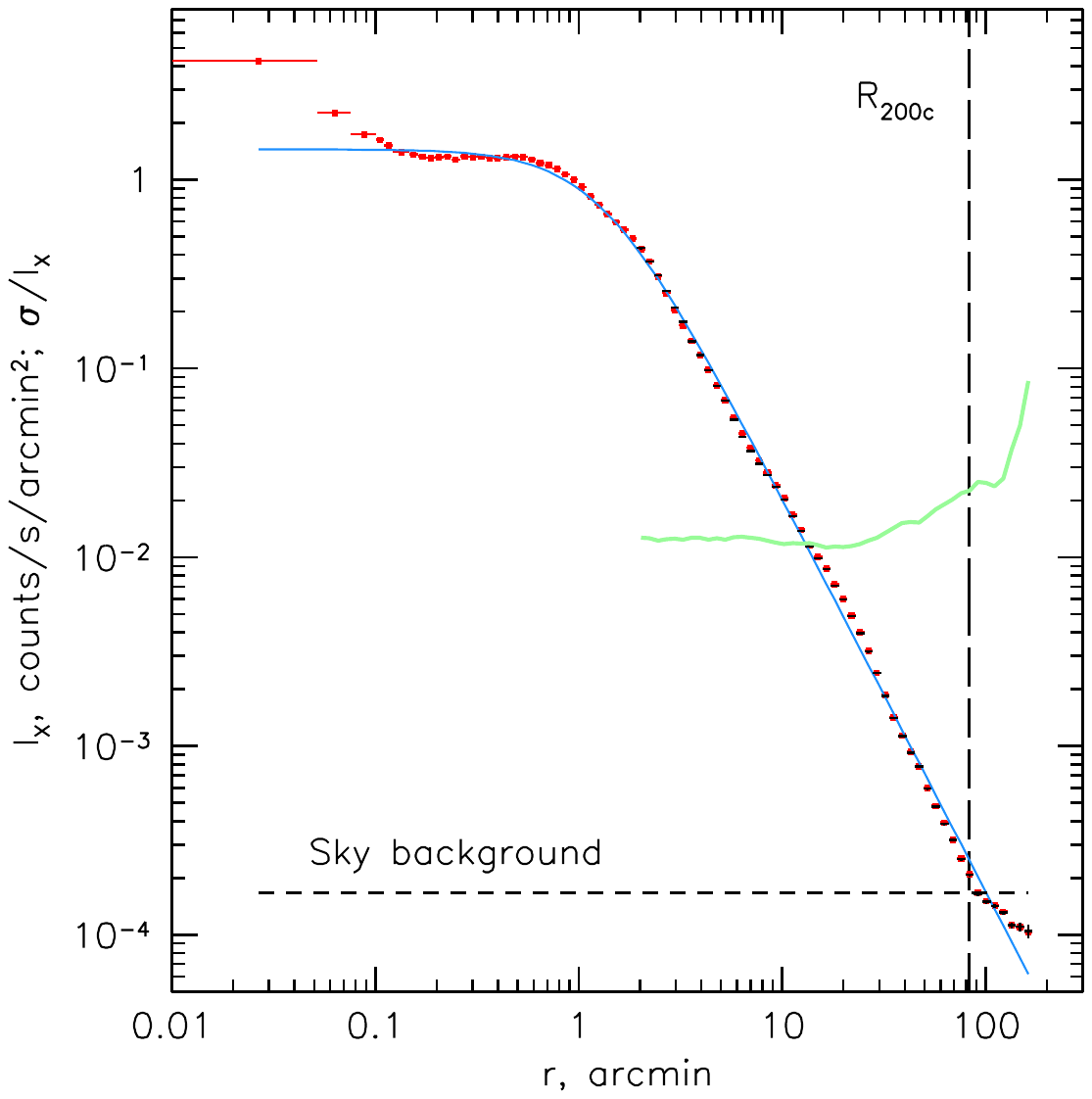}
\caption{Radial X-ray surface brightness profile (red points) based on the X-ray image shown in Fig.\ref{f:cxee}. The blue solid line shows a $\beta$-model with $r_c=1.3'$ and $\beta=0.514$. The best-fitting constant sky background level (shown with the dashed black line) has been subtracted from the data and the model. The error bars associated with the photon counting noise are smaller than the size of the plotted symbols. The black points with error bars show the eROSITA-only radial profile for $r>2'$. The ratio of errors to the Perseus surface brightness is plotted as a green line.
Near the virial radius, the statistical errors are at the $\sim 2$\% level of the observed surface brightness. At large radii ($\sim 100'$), the contribution from the telescope stray light is important \citep[see Appendix A in][]{2023A&A...670A.156C} and makes a non-negligible contribution to the observed X-ray surface brightness. 
} 
\label{f:radial}
 \end{figure}

A radial profile of the resulting image is shown in Fig.~\ref{f:radial}. The observed profile was approximated with a simple combination of a $\beta$-model (the blue line in Fig.~\ref{f:radial}) and a constant (the dashed line). The very core ($\sim5$") was excluded when fitting the profile.  This model (sum of the $\beta$-model and the constant) is used to remove a symmetric component from the X-ray image and express the residual deviations relative to the same model.

\section{Impact of photoelectric absorption in the Milky Way on the surface brightness variations.}
\label{a:nh}
\begin{figure*}
\includegraphics[angle=0,clip,width=2\columnwidth]{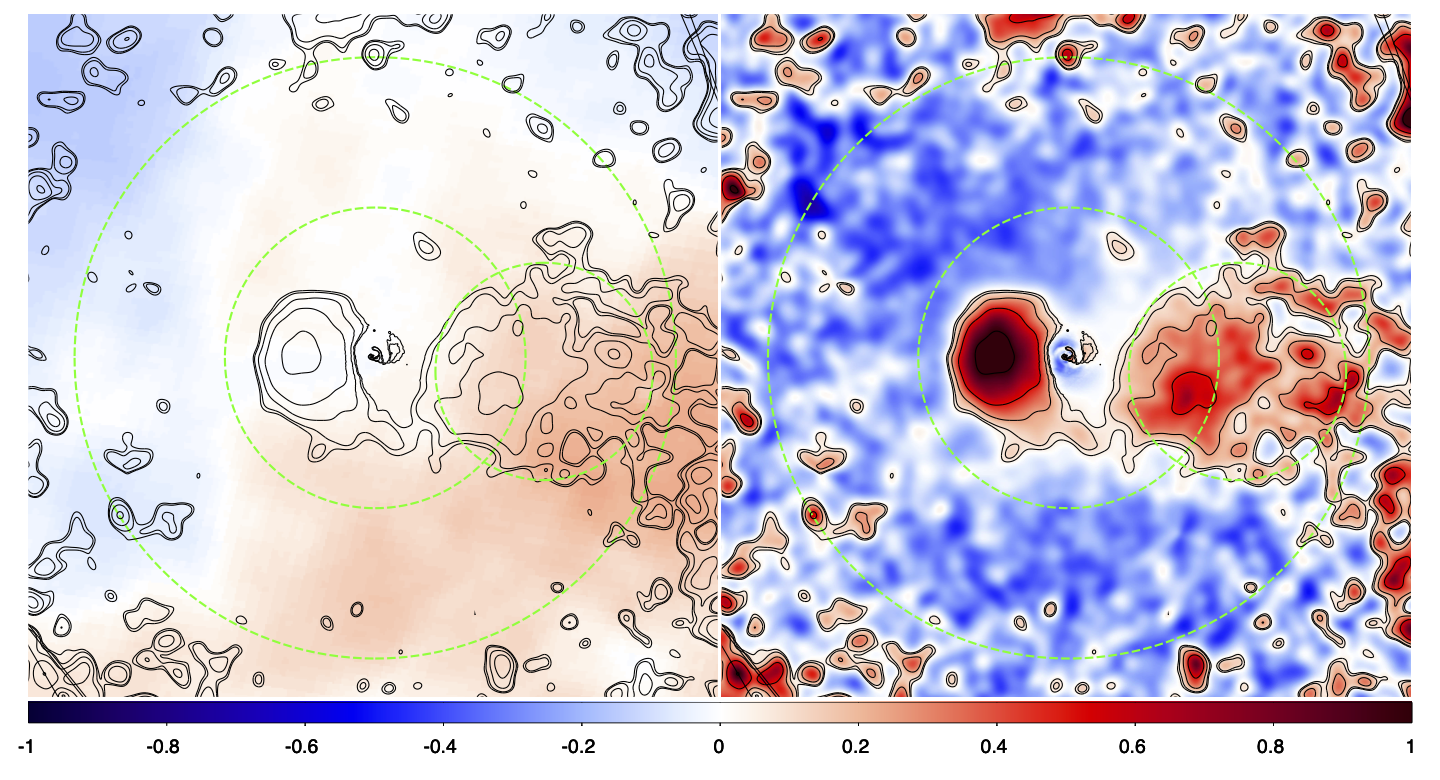}
\caption{Expected modulation of the X-ray flux (left panel) due to variation of the photoelectric absorption in the Galaxy in azimuthal direction (i.e., relative to the mean value at the same distance from NGC1275). The modulation was computed to 0.4-2.3 keV energy range, taking into account the eROSITA response. For comparison, the right panel shows the observed azimuthal variations of the X-ray flux in the same color scale. Clearly, the observed variations are several times larger than expected due to absorption. The two large dashed circles correspond to $0.5\times R_{200c}$ and $R_{200c}$. The smaller circle to the West of the core outlines the region of the most prominent patch of the excess X-ray emission, presumably caused by merger(s) along the largest filament crossing the Perseus cluster. The contours of the elevated X-ray surface brightness are shown in both panels for comparison.} 
\label{f:nh}
\end{figure*}

The center of the Perseus cluster is only $\sim 13$ degrees away from the Galactic Plane. This means that some of the observed variations of the X-ray surface brightness might be caused by spatially non-uniform photoelectric absorption of X-rays by neutral or molecular gas in the Galaxy. Here, we demonstrate that the amplitude of this effect is significantly smaller than the observed variations. To do so, we used the 3D dust maps from Bayestar19 \cite[see][for detailed description]{2019ApJ...887...93G}. We are looking for the intervening absorption 
of extragalactic X-rays from the cluster itself and from the Cosmic X-ray Background (CXB) in general. The total column density (NH) was calculated from the extinction maps based on the Bayestar19 data. The column density map was converted into an "attenuation map" of the X-ray flux in the 0.4-2.3 keV band, taking into account the eROSITA response. As a final step, the attenuation map in every pixel was divided by the mean value in concentric rings, mimicking the procedure used to generate Fig.~\ref{f:divc}. A direct comparison of the expected modulation with the X-ray data (Fig.~\ref{f:nh}) shows that the photoelectric absorption effect is subdominant. Furthermore, we have verified that in the harder 1-2.3 keV band, where photoelectric absorption is weaker, the X-ray image possesses similar large-scale structures as seen in Fig.~\ref{f:divc_200}.

\end{appendix}

\label{lastpage}
\end{document}